\documentclass[fleqn,usenatbib]{mnras}

\usepackage{newtxtext,newtxmath}
\usepackage[T1]{fontenc}
\usepackage{tablefootnote}

\DeclareRobustCommand{\VAN}[3]{#2}
\let\VANthebibliography\thebibliography
\def\thebibliography{\DeclareRobustCommand{\VAN}[3]{##3}\VANthebibliography}

\usepackage{graphicx}	% Including figure files
\usepackage{amsmath}	% Advanced maths commands
\usepackage{subfigure}

\title[ALMA REBELS survey: {[OIII]}$_{88\mu \text{m}}$ line scans]{The ALMA REBELS survey: [OIII]$_{88\mu \text{m}}$ line scans of UV-bright $z \gtrsim 7.6$ galaxies}

\author[I.F. van Leeuwen]{
I. F. van Leeuwen$^{1}$\thanks{E-mail: vleeuwen@strw.leidenuniv.nl},
R. J. Bouwens$^{1}$,
J. A. Hodge$^{1}$,
P. P. van der Werf$^{1}$,
H. S. B. Algera$^{2}$,
S. Schouws$^{1}$, \newauthor
M. Aravena$^{3,4}$,
R. A. A. Bowler$^{5}$,
P. Dayal$^{6}$,
A. Ferrara$^{7}$,
R. Fisher$^{5}$,
Y. Fudamoto$^{9}$,
C. Gulis$^{10}$, \newauthor
T. Herard-Demanche$^{1}$, 
H. Inami$^{11}$, 
I. de Looze$^{12}$,
A. Pallottini$^{7,8}$,
R. Smit$^{13}$,
L. Sommovigo$^{14}$,
and M. Stefanon$^{15,16}$
\\
$^{1}$Leiden Observatory, Leiden University, NL-2300 RA Leiden, the Netherlands\\
$^{2}$Institute of Astronomy and Astrophysics, Academia Sinica, 11F of Astronomy-Mathematics Building, No.1, Sec. 4, Roosevelt Rd, Taipei 106216, Taiwan, R.O.C.\\
$^{3}$Instituto de Estudios Astrof\'{\i}cos, Facultad de Ingenier\'{\i}a y Ciencias, Universidad Diego Portales, Av. Ej\'ercito 441, Santiago, Chile \\
$^{4}$Millenium Nucleus for Galaxies (MINGAL) \\
$^{5}$Jodrell Bank Centre for Astrophysics, Department of Physics and Astronomy, School of Natural Sciences, The University of Manchester, Manchester, M13 9PL, UK \\
$^{6}$Kapteyn Astronomical Institute, University of Groningen, P.O. Box 800, 9700 AV Groningen, The Netherlands \\
$^{7}$Scuola Normale Superiore, Piazza dei Cavalieri 7, 56126, Pisa, Italy\\
$^{8}$Dipartimento di Fisica ``Enrico Fermi'', Universit\'{a} di Pisa, Largo Bruno Pontecorvo 3, Pisa I-56127, Italy\\
$^{9}$Center for Frontier Science, Chiba University, 1-33 Yayoi-cho, Inage-ku, Chiba 263-8522, Japan\\
$^{10}$ Department of Astronomy, the Pennsylvania State University, 525 Davey Lab, University Park, PA 16802\\
$^{11}$Hiroshima Astrophysical Science Center, Hiroshima University, 1-3-1 Kagamiyama, Higashi-Hiroshima, Hiroshima 739-8526, Japan \\
$^{12}$Sterrenkundig Observatorium, Ghent University, Krijgslaan 281 S9, B-9000 Ghent, Belgium\\
$^{13}$Astrophysics Research Institute, Liverpool John Moores University, 146 Brownlow Hill, Liverpool L3 5RF, United Kingdom\\
$^{14}$Center for Computational Astrophysics, Flatiron Institute, 162 5th Avenue, New York, NY 10010, USA\\
$^{15}$Departament d’Astronomia i Astrofìsica, Universitat de València, C. Dr. Moliner 50, E-46100 Burjassot, València, Spain \\
$^{16}$Unidad Asociada CSIC ``Grupo de Astrofísica Extragaláctica y Cosmología" (Instituto de Física de Cantabria—Universitat de València), Spain \\
}

\date{Accepted XXX. Received YYY; in original form ZZZ}

\pubyear{2025}

\begin{document}
\label{firstpage}
\pagerange{\pageref{firstpage}--\pageref{lastpage}}
\maketitle

\begin{abstract}

\noindent We present the [OIII]$_{88\mu \text{m}}$ spectral scan results from the ALMA Large Program REBELS (Reionization Era Bright Emission Line Survey). The generally high luminosity of [OIII]$_{88\mu \text{m}}$ and ALMA’s Band 7 efficiency motivated its use for line scans of REBELS targets at $z>8$. Spectral scans of four sources covered 326.4–373.0 GHz ($z=8.10$–9.39), reaching [OIII]$_{88\mu \text{m}}$ luminosities of $\mathrm{\sim7.6\times10^8\ L_{\odot}}$ ($5\sigma$) for a FWHM of 400 km s$^{-1}$. No credible lines are detected for the four targets. For REBELS-04, the non-detection is unexpected given the $\geq92\%$ coverage of the redshift likelihood distribution and its estimated SFR of 40~$\text{M}_{\odot}\ \text{yr}^{-1}$. Possible explanations for the faint [OIII]$_{88\mu \text{m}}$ emission (assuming a FWHM of 100 km s$^{-1}$) include high ISM densities ($>n_{\text{crit}} \approx 510\ \text{cm}^{-3}$) and low ionization parameters ($\mathrm{log_{10}\ U_{ion}\lesssim -2.5}$).  For REBELS-37, a subsequent detection of [CII]$_{158\mu \text{m}}$ ($z=7.643$) confirmed it lay outside our scan range. For REBELS-11 and REBELS-13, it remains unclear if the non-detection is due to the depth of the line scan or redshift coverage. REBELS-04 and REBELS-37 show significant ($\geq3.8\sigma$) dust continuum emission in Band 7. If the photometric redshift of REBELS-04 is accurate, i.e., $z_{\mathrm{phot}}=8.57^{+0.10}_{-0.09}$ or $z_{\mathrm{phot}}=8.43^{+0.10}_{-0.10}$ accounting for additional neutral hydrogen in the circumgalactic medium, REBELS-04 would constitute the most distant dust-detected galaxy identified with ALMA to date. Additional Band 6 dust observations of REBELS-37 constrain the shape of the far-IR SED, ruling out cold dust temperatures ($\lesssim28$ K) at $3\sigma$. Further insight into these galaxies will require spectroscopic redshifts and deeper multi-band dust observations. 
\end{abstract}


%%%%%%%%%%%%%%%%%%%%%%%%%%%%%%%%%%%%%%%%%%%%%%%%%%

%%%%%%%%%%%%%%%%% BODY OF PAPER %%%%%%%%%%%%%%%%%%

\section{Introduction}
The Epoch of Reionization (EoR) is a prominent era in the history of our Universe. In the EoR the first galaxies start to reionize the predominantly neutral Universe \citep[e.g.][]{dayal2018}. The galaxies in the EoR are expected to differ significantly from the present day galaxies, as these galaxies have had less time to enrich their ISM with metals and dust and are expected to generate harder ionizing emission (e.g. \citealt{katz2020,Tang2023}). Fortunately, great leaps in instrumentation and new telescopes (e.g. \textit{JWST}) have allowed us to peer into this early era of the Universe and we can start to characterize the galaxies in the EoR \citep[e.g.][]{Robertson2022, adamo2024}.

To understand the formation and evolution of galaxies in the EoR, the use of far-infrared (FIR) emission lines is particularly useful. FIR lines are expected to be unaffected by dust extinction and some are exceptionally bright and therefore useful for observing distant galaxies. Two notably bright FIR lines are [CII]$_{158\mu\text{m}}$ and [OIII]$_{88\mu\text{m}}$ that predominantly trace the photo-dissociation \citep[e.g.][]{Wolfire2003} and ionized regions \citep[e.g.][]{Cormier2015}, respectively. 

The Atacama Large Millimeter/submillimeter Array (ALMA) has been instrumental in the exploration of the high redshift Universe with FIR emission lines \citep{hodge_cunha}. The [CII]$_{158\mu\text{m}}$ emission line has been extensively used to detect and characterize galaxies and quasars at $z \geq 4$ (e.g. \citealt{capak2015, bethermin2020, fevre2020, venemans2020}), while at higher redshifts ($z > 8$) the [OIII]$_{88\mu\text{m}}$ line is predicted to be more efficient \citep{inoue2014}. Local low-metallicity dwarfs have been found to be brighter in [OIII]$_{88\mu\text{m}}$ than in [CII]$_{158\mu\text{m}}$  \citep{cormier2012, madden2013} which suggest that [OIII]$_{88\mu\text{m}}$  could be a more efficient line to target when observing high redshift and potentially low-metallicity galaxies. 
 
The first detection of [OIII]$_{88\mu\text{m}}$ at $z > 7$ was achieved in a Ly-$\alpha$ emitter \citep{inoue2016}. The [OIII]$_{88\mu\text{m}}$ emission line has become a preferred FIR line at $z > 8$ to derive spectroscopic redshifts for galaxies and characterize them in detail (e.g. \citealt{hashimoto2018, tamura2019, fujimoto2024}). Recent observations of two of the most distant spectroscopically confirmed galaxies, JADES-GS-z14-0 and GHZ2, have even proved that ALMA is capable of detecting [OIII]$_{88\mu\text{m}}$ line emission at $z > 12$ \citep{schouws2024, carniani2024, zavala2024}.  

The [OIII]$_{88\mu\text{m}}$ emission line is thought to arise mainly from HII regions \citep[e.g.][]{Cormier2015}, due to its high ionization potential ($E > 35.1$ eV, \citealt{malhotra2001}). Not only is the [OIII]$_{88\mu\text{m}}$ emission line utilized in spectral scans for redshift determinations, this FIR line can also characterize the ISM. The brightness of [OIII]$_{88\mu\text{m}}$ for a given SFR increases with increasing ionization parameter ($\text{U}_{\text{ion}}$), lower hydrogen densities ($n_{\text{H}}$) than its critical density ($n_{\text{crit}} \approx 510\ \text{cm}^{-3}$) and increasing metallicity ($Z$) (e.g. \citealt{moriwaki2018, harikane2020, Kohandel2023, nakazato2023}).

The Reionization Era Bright Emission Line Survey (REBELS) is an ALMA Large Program designed to construct a sample of especially massive ISM reservoirs in the Reionization Era ($z > 6.5$: \citealt{REBELS}). To achieve this goal, REBELS performed an extensive line search for luminous [CII]$_{158\mu\text{m}}$ (36 targets) or [OIII]$_{88\mu\text{m}}$ (4 targets) line emission from 40 UV-selected galaxies, while also probing the dust continuum. The REBELS observations started in Cycle 7 and were completed in Cycle 8. A total of 25 galaxies have been detected in [CII]$_{158\mu\text{m}}$ and are presented in \citet{REBELS} and Schouws et al. in preparation. The REBELS sample has enabled the study of properties of a substantial number of EoR galaxies, such as dust continuum emission \citep{Inami2022}, dust properties like attenuation and temperature \citep{dayal2022, ferrara2022, sommovigo2022, bowler2024, algera_oiii_cii, palla2024}, obscured star formation rate density \citep{algera23} and the IR luminosity function at $z \sim 7$ \citep{barrufet2023}. Moreover, the REBELS survey enabled investigation into specific SFRs \citep{topping2022}, serendipitously detected dusty galaxies \citep{fudamoto2021}, [CII]$_{158\mu \text{m}}$ sizes \citep{fudamoto2022}, Lyman-$\alpha$ emission \citep{endsley2022}, HI gas masses \citep{heintz2022} and molecular gas \citep{aravena2024} in the EoR and resolved studies of the most massive galaxy of the sample (REBELS-25; \citealt{hygate2023, rowland2024}).

In this paper, we present the [OIII]$_{88\mu\text{m}}$ line scans performed for four of the highest redshift ($z > 7.6$) galaxies of the REBELS sample in both ALMA Cycle 7 and 8.  We also present an updated assessment of the dust continuum emission from the [OIII]$_{88\mu\text{m}}$ targeted sources, refining the previous measurements from \citet{Inami2022} using Cycle 7 data only.  In addition, for one of our targets (REBELS-37), Band 6 dust continuum observations are now available, which can be combined with the constraints from Band 7 measurements to probe both the dust temperature and shape of the far-IR continuum.

This work is organized as follows. In Section \ref{sec:obs_reduction}, we introduce the [OIII]$_{88\mu\text{m}}$ observations of REBELS and the data reduction. Additionally we present the coverage of the redshift likelihood distribution of the $z \gtrsim 7.6$ sources and discuss the procedures we use to measure [OIII]$_{88\mu\text{m}}$ and IR luminosity. In Section \ref{sec:results}, we present the [OIII]$_{88\mu\text{m}}$ line scan data and significant Band 7 dust continuum detections. We also examine the dust temperature of REBELS-37. We discuss our interpretation of the [OIII]$_{88\mu\text{m}}$ line scan results as well as improved constraints on the dust continuum and dust temperature for our sample in Section \ref{sec:discussion}. Finally, our findings are summarized in Section \ref{sec:conclusions}.  
In this work we adopt a flat $\Lambda$CDM cosmology with h~=~0.7, $\Omega_{\text{M}}$ = 0.3, and $\Omega_{\Lambda}$ = 0.7. 

\section{Observations and data reduction}
\label{sec:obs_reduction}
\subsection{The ALMA REBELS survey} 
The ALMA Large Program REBELS (2019.1.01634.L, PI: Bouwens) targeted 40 UV-luminous ($ -23.0 <\text{M}_{\text{UV, AB}} < -21.3$) star-forming galaxies at $z > 6.5$ \citep{REBELS}. REBELS was designed as a spectral scan program to target [CII]$_{158 \mu \text{m}}$ (Bands 5 and 6) or [OIII]$_{88 \mu \text{m}}$ (Band 7) emission, while simultaneously obtaining dust continuum measurements. REBELS observations were mainly performed in Cycle 7, except for the observations of six targets that were completed in Cycle 8. The targets for the REBELS survey were selected over a 7 deg$^2$ area to have SFR$_{\text{UV}}$ $> 11\ \text{M}_{\odot}\ \text{yr}^{-1}$ and to have relatively tight constraints on their photometric redshift to allow for efficient line scans.\footnote{For more details on the sample selection and observing strategy see \citet{REBELS}.} The [CII]$_{158\mu\text{m}}$ line scans and detections are discussed in \citet{REBELS} and Schouws et al. (in preparation).  Initial (Cycle 7) results from REBELS on the dust-continuum emission and luminosity of sources are presented in \citet{Inami2022}. In this work we present Cycle 7+8 results from [OIII]$_{88\mu\text{m}}$ line scans performed on four $z\gtrsim 7.6$ targets from the REBELS program: REBELS-04, REBELS-11, REBELS-13 and REBELS-37.

\subsection{Data reduction}
We have calibrated the ALMA measurement sets using the ScriptforPI with the Common Astronomy Software Applications (\textsc{casa}) software \citep{CASA}. The calibrated measurement sets of both Cycle 7 and 8 are combined and time-averaged over 30 s (see Schouws et al. in preparation). The observations are imaged with \textsc{casa} version 6.5.4. The data is cleaned with the \textsc{tclean} task to a depth of 2$\sigma$ using `auto-multithresh' \citep{multithresh} applying Briggs weighting with the robust parameter set to 2.0 (therefore resembling natural weighting). The pixel sizes are set such that there are approximately 5 pixels along the minor axis of the beam. Unfortunately for all sources one spectral window (SPW) of tuning 3 (covering the redshift range $z=8.188$-8.240) is affected by an atmospheric line at 368.5 GHz (O$_2$ N$_J$ = 3$_2$ $\rightarrow$ 1$_1$) and caused the transmission to be down to $\sim 0\%$ and therefore unusable. We have therefore excluded this SPW during the analysis. The \textsc{casa} task \textit{uvcontsub} is used for subtracting any continuum emission from potential [OIII]$_{88\mu\text{m}}$ line emission by a fitting zeroth order polynomial. As will be discussed in the following sections, we did not find any [OIII]$_{88\mu\text{m}}$ emission lines in the data, but we did find significant ($\geq$ 3.8$\sigma$) continuum emission. The properties of the resulting image cubes are summarized in Table~\ref{tab:observations}.

\begin{table*}
\centering
\begin{tabular}{ccccccccc}
\hline
REBELS ID & Source name & RA & Dec & Beam & Noise$_{\text{cube}}$ & $\mathrm{\nu_{cont, obs}}$& Noise$_{\text{cont}}$ \\ 
 & & & & (arcsec$^{2}$) & (mJy beam$^{-1}$ channel$^{-1}$) & (GHz) & ($\mu$Jy beam$^{-1}$) \\ \hline
 \multicolumn{8}{|c|}{\textbf{Band 7}} \\ 
REBELS-04 & XMM-J-355 & 02:17:42.46 & 
-04:58:57.4 & 0.97 $\times$ 0.82 & 0.46 & 351.52 & 18.0 \\
REBELS-11 & XMM3-Y-217016 & 02:24:39.35 & 
-04:48:30.0 & 0.97 $\times$ 0.82 & 0.47 & 351.53 & 24.4 \\
REBELS-13 & XMM-J-6787 & 02:26:16.52 & 
-04:07:04.1& 0.97 $\times$ 0.80 & 0.68 & 349.67& 22.0 \\
REBELS-37 & UVISTA-J-1212 & 10:02:31.81 & 
+02:31:17.1 &  0.96 $\times$ 0.85 & 0.42 & 349.66& 15.8 \\

 \multicolumn{8}{|c|}{\textbf{Band 6}} \\ 
 REBELS-37 & UVISTA-J-1212 & 10:02:31.81 & 
+02:31:17.1 &  $1.69 \times 1.41^{*}$ & - & 219.16& 12.6 \\
 \hline
\end{tabular}
\caption{Summary of the ALMA Band 7 observations with the target IDs, central positions, median beam size of the image cube, median noise in the image cube per $\sim$13 km s$^{-1}$ channel, the observed frequency of the continuum image and noise in the continuum image estimated from the dirty continuum image. UVISTA-J-1212 also has observations in Band 6 (Schouws et al., in preparation) and we additionally analyze the Band 6 continuum in this work.\\
$^{*}$ The beam size of the Band 6 data of UVISTA-J-1212 presented here is of the continuum image.} 
\label{tab:observations}
\end{table*}

\subsection{Coverage of redshift likelihood distribution}
\label{sec:coverage_z}
\begin{figure*}
    \centering
    \begin{subfigure}
        \centering
        \includegraphics[width=0.475\textwidth]{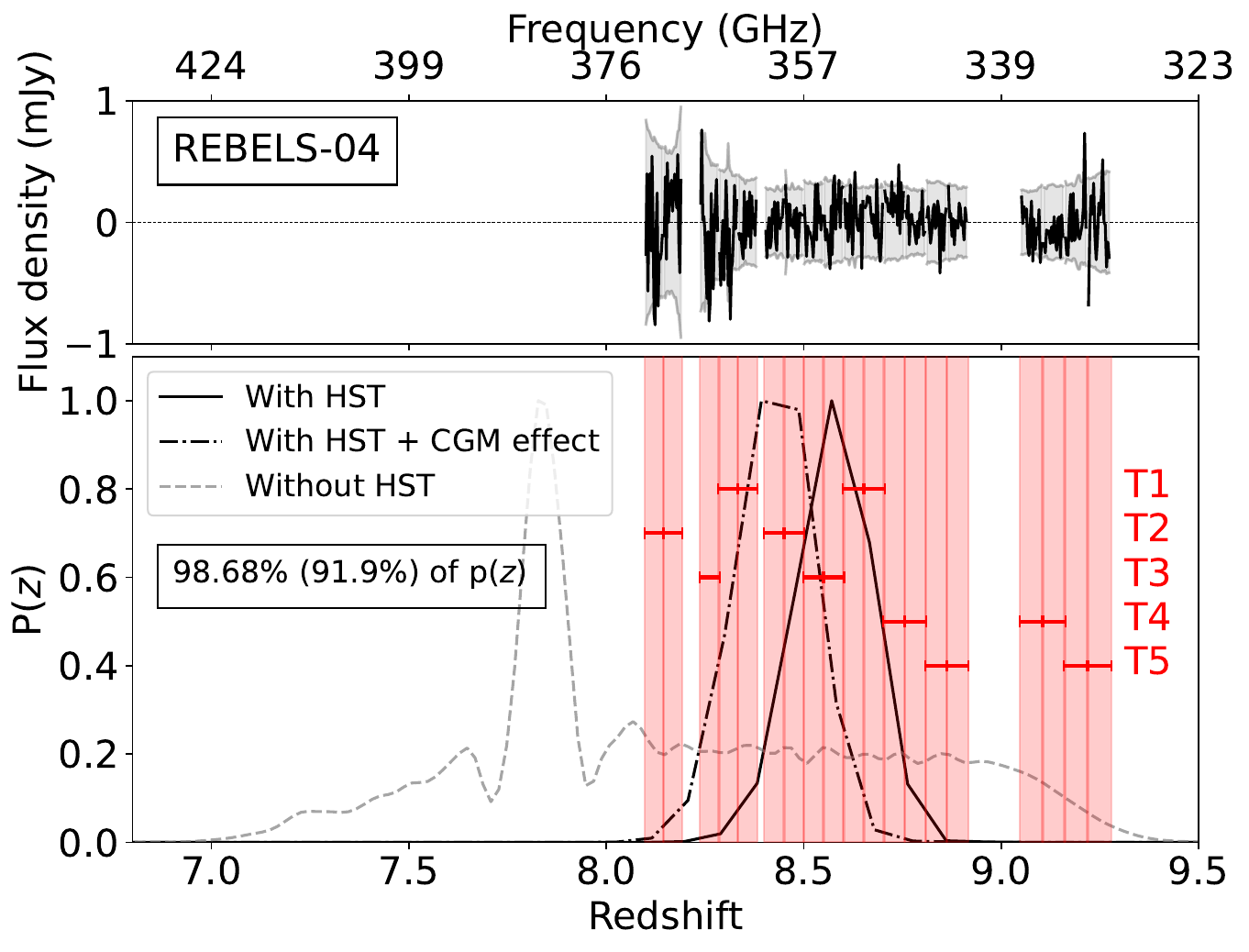}
    \end{subfigure}
    \hfill
    \begin{subfigure}
        \centering
        \includegraphics[width=0.475\textwidth]{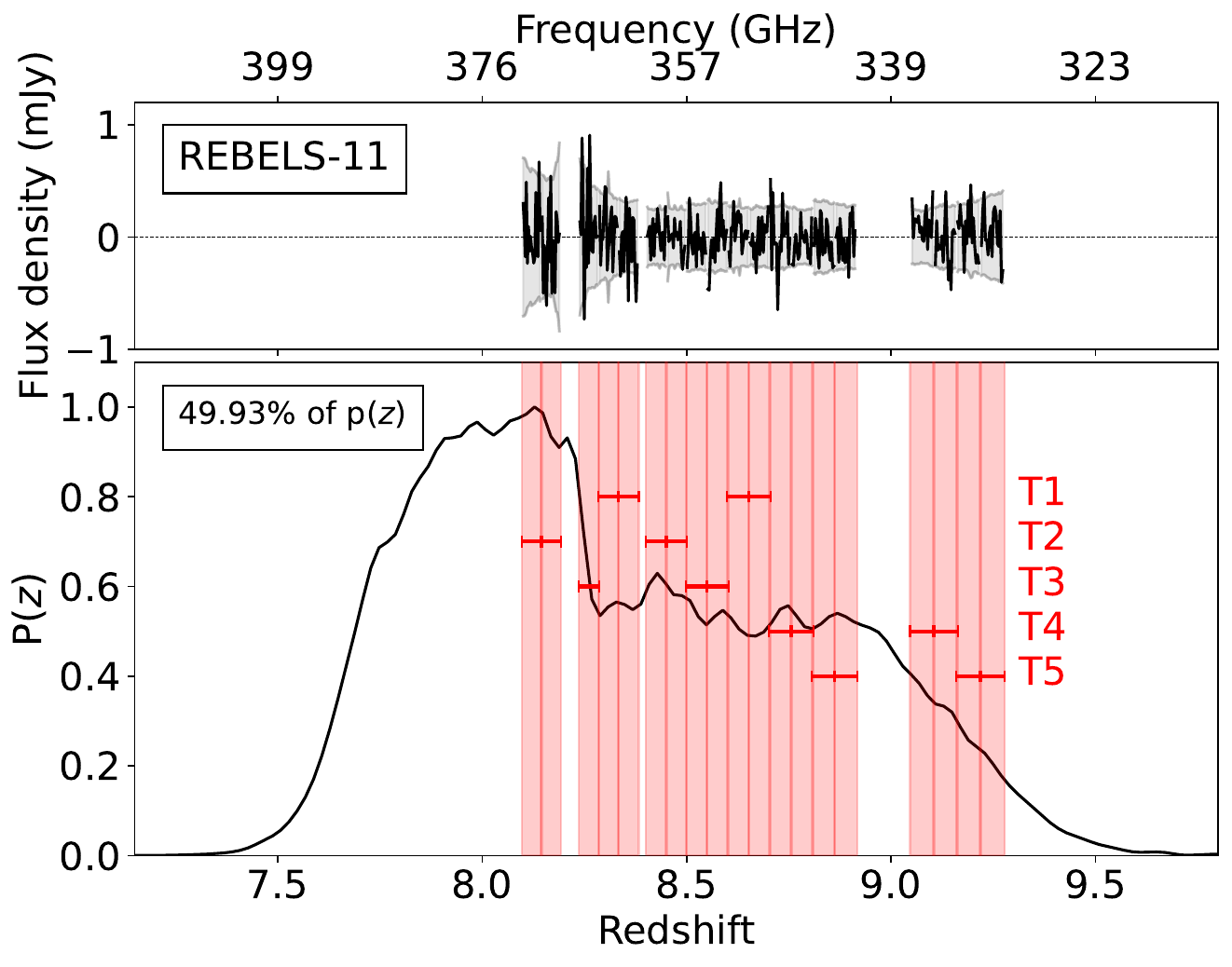}
    \end{subfigure}
    \vskip\baselineskip
    \begin{subfigure}
        \centering
        \includegraphics[width=0.475\textwidth]{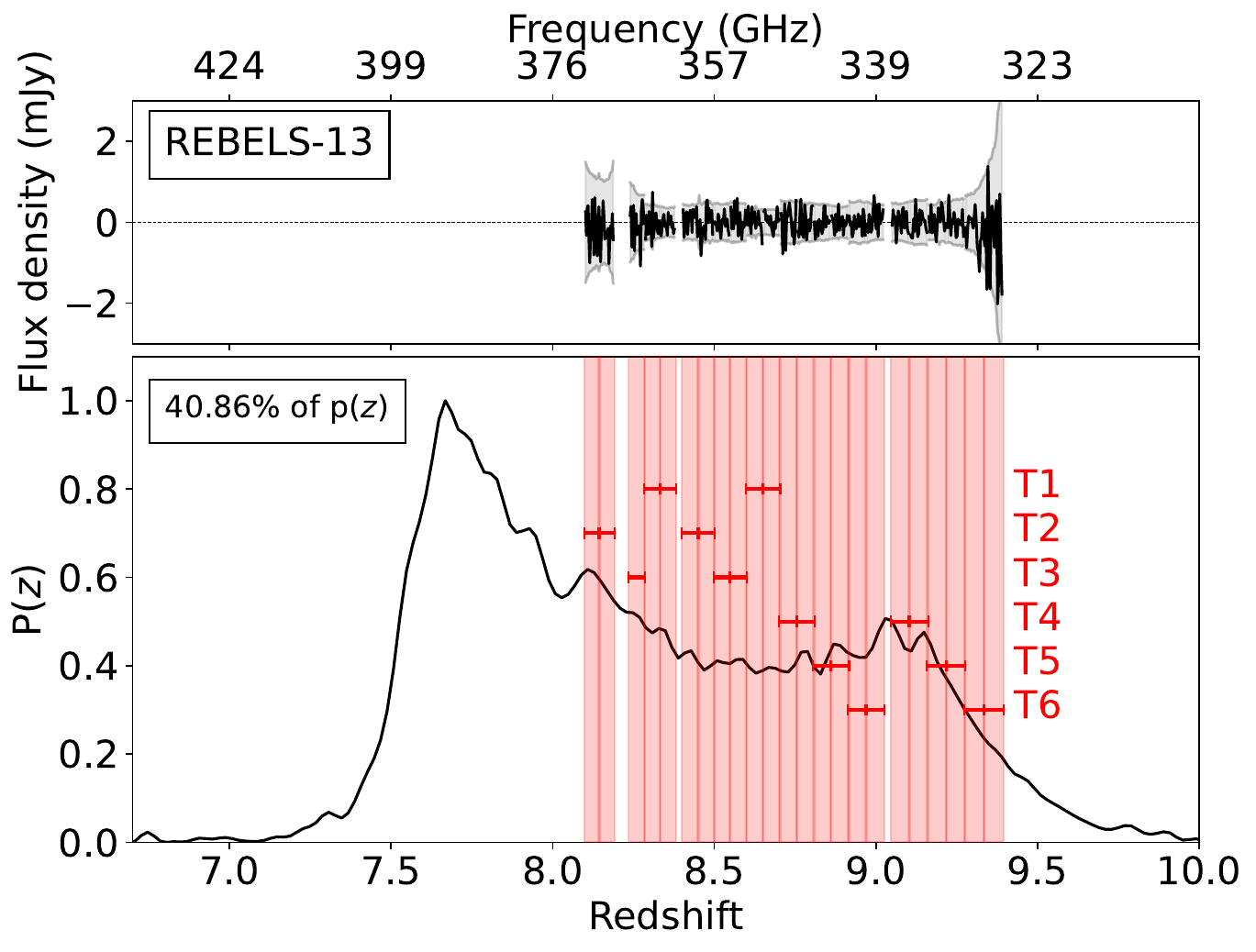}
    \end{subfigure}
    \hfill
    \begin{subfigure}
        \centering
        \includegraphics[width=0.475\textwidth]{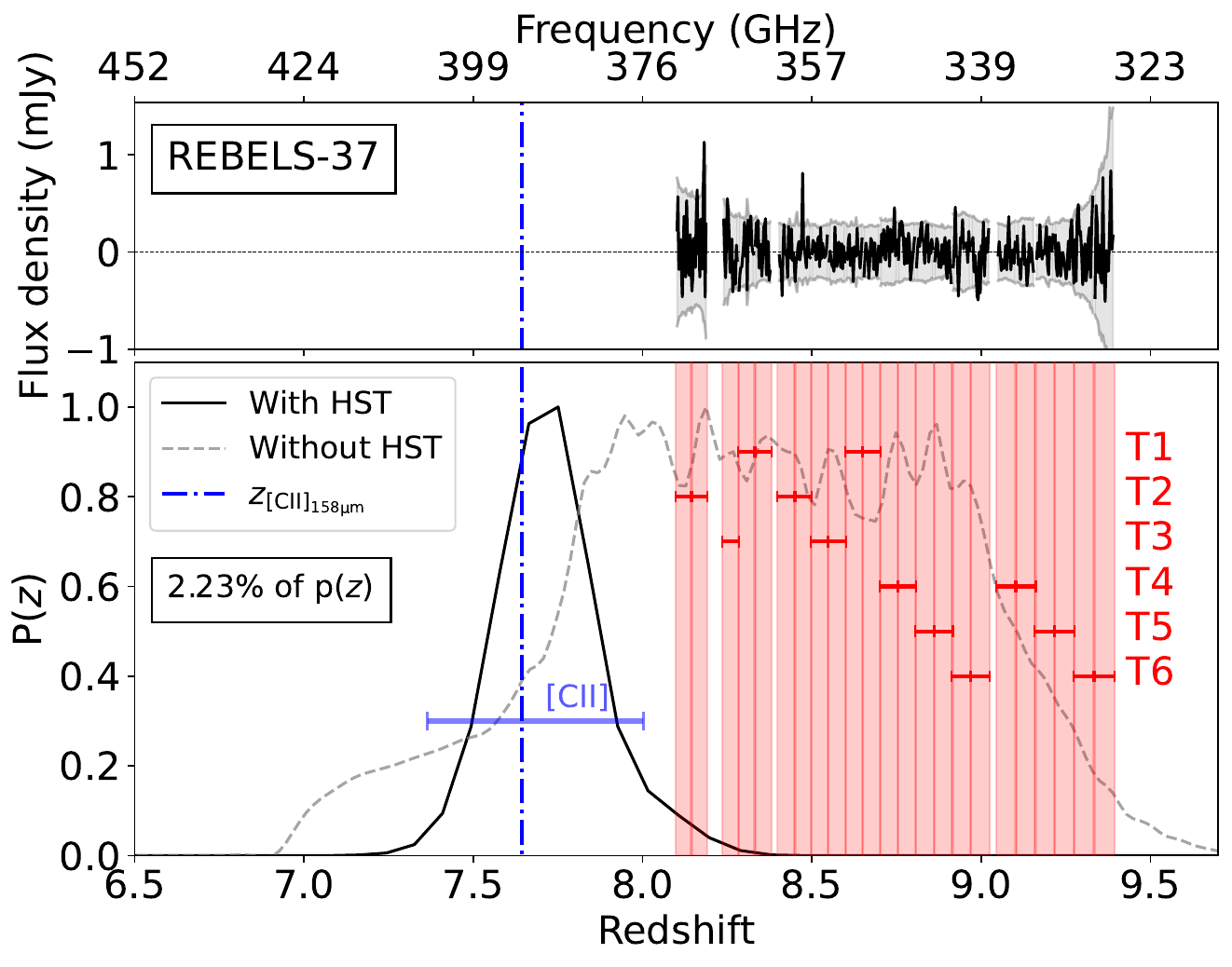}
    \end{subfigure}
    \caption{\textit{Top rows}: Spectral extractions show our [OIII]$_{88\mu\text{m}}$ line scans, binned in intervals of 90 km s$^{-1}$. The spectra are extracted by taking a circular aperture with an area equal to the beam size centered on rest-UV position of each target. The 1$\sigma$ uncertainty is calculated at each frequency by taking 1000 random apertures outside of the center of the image and measuring the standard deviation.
    \textit{Bottom rows:} The redshift likelihood distributions of the sources based on both ground-based and space observations. The tighter redshift constraints available for REBELS-04 and REBELS-37 are due to additional \textit{HST} F105W, F125W, and F160W observations.  For context, the weaker redshift constraints available for these sources while initially setting up the scan strategy are shown with dashed lines. Additionally, for REBELS-04 the redshift distribution is shown when an additional CGM component with neutral hydrogen is added to the SED fit. The 5 to 6 tunings (T1-T6) of the SPWs of ALMA Band 7 are used in an effort to cover the redshift likelihood distributions. There is a gap around 368 GHz due to an atmospheric line that reduces the transmission to 0$\%$ in this SPW. For REBELS-37 we also indicate the redshift range covered by the SPWs used to target [CII]$_{158\mu\text{m}}$ in Cycle 8 and the spectroscopic redshift from the line detection together with the frequency of the [OIII]$_{88\mu\text{m}}$ line this would correspond to (presented in Schouws et al. in preparation: see also \citealt{harikane2024}).}
    \label{fig:pz_spws}
\end{figure*}

The REBELS survey was designed to efficiently cover the redshift likelihood distribution of the 40 UV-luminous sources with the SPWs of ALMA Band 5, 6 and 7. The redshift likelihood distributions are derived using several independent methods and are explained in \citet{REBELS}. Most of the sources with photometric redshifts larger than 7.6 are targeted with the [OIII]$_{88\mu \text{m}}$ line and have larger uncertainties on their redshift due to the fact that the Lyman break is between the ground-based Y and J bands. Additional Hubble Space Telescope (\textit{HST}) F105W, F125W, and F160W observations of REBELS-04 and REBELS-37 (GO 16879, PI: Stefanon and GO 15931, PI: Bowler) were therefore instrumental in substantially improving the constraints on the redshifts of these sources. The \textit{HST} observations were executed after ALMA had already started the observations for the REBELS program. The redshift likelihood distributions (P($z$)) of the four sources discussed in this work are shown in the bottom panels of Fig.~\ref{fig:pz_spws} by the black lines. The gray dashed lines are the redshift likelihood  distributions prior to the additional \textit{HST} data. For all sources, five to six tunings with four SPWs (shown in red) were used to cover a significant portion of the redshift distributions.\footnote{There is a gap in the frequencies covered around $\sim368.5$ GHz in the spectra due to exclusion of the SPW that covers an atmospheric line resulting in 0$\%$ transmission.}

The two most tightly constrained redshift likelihood  distributions are for those sources targeted with additional \textit{HST} imaging. The photometric redshift of REBELS-04, in particular, is very well determined due to this additional \textit{HST} imaging to constrain the precise wavelength of the Lyman break, but also thanks to the IRAC photometry available of the source, which show a relatively flat UV continuum from 3.6$\mu$m to 1.3$\mu$m.  The 4.5$\mu$m flux shows a $\sim$0.7-mag excess over that present in $UV$ continuum, likely due to luminous [OIII]$_{4959,5007}$ emission.  Together with strong evidence for the existence of a Lyman break in the source, the existence of a second feature in the spectral energy distribution (SED) should make the photometric redshift of REBELS-04 very robust. 

Recent work by \citet{asada2024} has discussed the additional impact that hydrogen clouds in the circumgalactic medium (CGM) can have on the derived photometric redshifts when added to the standard intergalactic medium (IGM) absorption model. \citet{asada2024} find that the photometric redshift can be overestimated by $\delta z = 0.20$ when only accounting for the standard IGM absorption. In Appendix~\ref{sec:phot_redshift} we show the photometry and SED of REBELS-04. Additionally, we examine the SED and redshift likelihood of REBELS-04 when an additional CGM component is added to the analysis and find a decrease in the derived photometric redshift ($z_{\mathrm{phot}} = 8.43^{+0.10}_{-0.10}$ instead of $z_{\mathrm{phot}} = 8.57^{+0.10}_{-0.09}$). The coverage of the redshift distribution is slightly smaller at 92$\%$ instead of $99 \%$ when we include the effect of the CGM.

The majority ($\geq 92\%$) of the redshift distribution has been scanned for REBELS-04, making it extremely likely that the [OIII]$_{88\mu\text{m}}$ line should be within the observed frequencies. In contrast,  REBELS-37 only has coverage of $2.2\%$ of the redshift  distribution in Band 7. This is due to the large change in the redshift likelihood distribution that came from the addition of new \textit{HST} observations midway through the execution of the program and after the Band 7 line scan had already been performed.  The \textit{HST} data shifted the peak of the redshift distribution to $z \lesssim 8$ and therefore for the Cycle 8 observations a request was made to finish the line scan using the [CII]$_{158\mu\text{m}}$ line. The [CII]$_{158\mu\text{m}}$ line has been detected in the Band 6 data with $\sim$10$\sigma$ (Schouws et al. in preparation: see also \citealt{harikane2024}), therefore measuring the spectroscopic redshift ($z_{\text{[CII]}}$~=~7.643). 
Unfortunately, the redshifted [OIII]$_{88\mu\text{m}}$ frequency implied by the spectroscopic redshift is outside of the SPWs. Therefore we cannot provide an upper limit on [OIII]$_{88\mu\text{m}}$ for REBELS-37. 

The other two sources, REBELS-11 and REBELS-13, have less tightly constrained redshift likelihood distributions.  Because of this, scanning for ISM cooling lines would have required covering the entire redshift range from $z\sim$ 7.6 to 9.4. Unfortunately, there was an insufficient amount of observing time available to REBELS to span the full range and were only able to cover the $z>$ 8.1 portion.  This results in a coverage of 49.9$\%$ and 40.9$\%$ of the redshift likelihood  distribution for REBELS-11 and REBELS-13, respectively.  We furthermore caution, as noted in \citet{REBELS}, that REBELS-13 could be a lower-redshift interloper given that $>$15$\%$ of the redshift distribution extends to $z < 6$.  With these caveats in mind, we provide tentative upper limits on the [OIII]$_{88\mu\text{m}}$ fluxes of these sources.

\subsection{[OIII]$_{88\mu\text{m}}$ line emission}
\label{sec:method_oiii}
To identify [OIII]$_{88\mu \text{m}}$ line emission in the datacubes we used Matched Filtering in 3D (\textsc{mf3d}; \citealt{mf3d}). \textsc{mf3d} uses 1D Gaussian frequency templates and 2D Gaussian spatial templates to convolve with the ALMA cubes to find line emission peaks. We adopt frequency templates with widths from 50 to 800 km s$^{-1}$ and spatial templates from 0 to 3 $\times$ the beam size. Schouws et al. in preparation has done extensive checks on the fidelity of REBELS sources as a function of SNR identified by \textsc{mf3d} (SNR$_{\text{MF3D}}$). Schouws et al. in preparation find a purity $> 95\%$ for emission peaks with SNR$_{\text{MF3D}}$ $\geq 5.2$ within a 1.5" radius of the center and SNR$_{\text{MF3D}}$ $\geq 6.2$ for peaks outside of the image center. SNR$_{\text{MF3D}}$ is based on the optimally-weighted signal when close to the provided template shape. We do not find any positive peaks within 1.5 arcsec radius from the center of the observations (which are centered to the UV-emission), except for a SNR$_{\text{MF3D}} = 5.35$ peak 1.45" from the center of the REBELS-13 observations. We note however that for the same observations more significant noise (negative) peaks are found (SNR$_{\text{MF3D}} = 5.9$) and as this emission is significantly offset from the UV emission, it is likely that this is a noise peak. Further analysis on this potential noise peak is shown in Appendix~\ref{sec:6787}. We do not find any spurious line emission peaks with SNR $\geq 6.2$ 1.5" outside of the center. We assume that the sources are unresolved, as we expect [OIII]$_{88\mu\text{m}}$ to generally be more compact than our spatial resolution (e.g. \citealt{tamura2019,witstok22, ren2023}). We extract spectra of the central region with circular apertures with areas equal to the beam and we cannot identify any line emission as well, see Figure ~\ref{fig:pz_spws}. 

With the exception of REBELS-37, we can provide $5\sigma$ upper limits on the [OIII]$_{88\mu\text{m}}$ fluxes. The full width at half maximum (FWHM) of the [OIII]$_{88\mu\text{m}}$ lines are unknown in this case. To calculate the [OIII]$_{88\mu\text{m}}$ flux, we therefore adopt two different values for the FWHM: 100 and 400 km $\text{s}^{-1}$. The choice of these FWHMs is motivated by $z>7$ [OIII]$_{88\mu\text{m}}$ line detections with FWHM $\sim 100$ km $\text{s}^{-1}$ (SXDF-NB1006-2 \citep{inoue2016}, MACS0416$\_$Y1 \citep{tamura2019} and MACS1149-JD1 \citep{hashimoto2018}. At $z>7$ more extreme values for the [OIII]$_{88\mu\text{m}}$ FWHM are also found that could be more representative for merging systems, for example in sources B14-65666 and A1689-zD1 ($>$ 300 km $\text{s}^{-1}$) \citep{hashimoto2019, wong2022}. The values of 100 and 400 km $\text{s}^{-1}$ are therefore representative of observations and have been used in other work discussing [OIII]$_{88\mu\text{m}}$ non-detections (e.g. \citealt{binggeli2021, popping2023}). 

We simulate moment-0 maps by collapsing the data over 2~$\times$~FWHM at the frequencies covered by the SPWs. The noise is estimated by taking 1000 random apertures with the size of the beam and then taking the standard deviation of flux densities. Using the $5\sigma$ noise on the flux density we can obtain an upper limit on [OIII]$_{88\mu\text{m}}$ luminosity flux using the following relation:

\begin{equation}
   L_{\text{[OIII]$_{88\mu\text{m}}$}} / \text{L}_{\odot} = 1.04 \times 10^{-3} \ \left(\frac{\nu_{\text{[OIII]}_{88\mu \text{m}, \text{obs}}}}{\text{GHz}}\right) \ \left(\frac{F_{\text{[OIII]$_{88\mu\text{m}}$}}}{\text{Jy km s}^{-1}}\right)\  \left(\frac{D_{L}}{\text{Mpc}}\right)^2.
   \label{eq:loiii}
\end{equation}

\noindent $F_{\text{[OIII]$_{88\mu\text{m}}$}}$ is the velocity integrated flux in $\text{Jy km s}^{-1}$ for which we use the $5\sigma$ upper limit, $\nu_{\text{[OIII]}_{88\mu \text{m}, \text{obs}}}$ the central frequency of the [OIII]$_{88\mu\text{m}}$ emission in GHz and $D_L$ the luminosity distance in Mpc \citep{solomon}.

\subsection{Dust continuum measurements}
\label{sec:method_dust}
We create the continuum maps with the \textsc{tclean} task of \textsc{casa} using \textit{specmode=`mfs'}. In the case of REBELS-37 we exclude the frequencies within 2 $\times$ the FWHM of the [CII]$_{158\mu\text{m}}$ line. To identify dust emission in the continuum maps we adopt the procedure from \citet{Inami2022} using \textsc{PyBDSF} \citep{pybdsf}. We summarize the procedure here. \textsc{PyBDSF} identifies neighboring emission (`islands') that are fitted with Gaussians to identify sources of emission. To establish the bounds of an island of emission and the peak of a source \textsc{PyBDSF} uses two SNR thresholds: iSNR and pSNR. \citet{Inami2022} extensively tested the purity of continuum sources with different iSNR and pSNR thresholds and find a purity of 95$\%$ for (iSNR, pSNR) = (2.0, 3.3). 

We find high-fidelity ((iSNR, pSNR) $\geq$ (2.0, 3.3)) Band 7 dust continuum emission for REBELS-04 and REBELS-37, consistent with the Cycle 7 data analyzed in \citet{Inami2022}. In this work we additionally make use of data we obtained in Cycle 8 for REBELS-04 and REBELS-11.  Thus, apart from REBELS-13 and REBELS-37 (where the full Band 7 data were obtained in Cycle 7), the present probe of the dust continuum flux is deeper than was available in \citet{Inami2022}. We use the peaks identified by \textsc{PyBDSF} to measure the continuum fluxes with the \textit{imfit} task from \textsc{casa}. We let \textit{imfit} fit a Gaussian in a region of 3" $\times$ 3" and obtain the continuum fluxes from the peak intensities. We use the peak intensities as \citet{Inami2022} found that all 40 REBELS sources, except for REBELS-25, are unresolved at 95$\%$ probability. 

We place a $3\sigma$ upper limit on the continuum fluxes of REBELS-11 and REBELS-13 using the noise in the dirty images, multiplied by a flux boosting correction of 1.33 (see \citealt{Inami2022}). Due to the change in the redshift likelihood distribution of REBELS-37, [CII]$_{158\mu\text{m}}$ was targeted for the remaining Cycle 8 observations in Band 6. We do not identify continuum emission in the Band 6 observations and therefore provide an upper limit. The resulting Band 7 (and Band 6 for REBELS-37) dust continuum fluxes are reported in Table~\ref{tab:sources} (uncorrected for the CMB effect). We note that with the additional Cycle 8 data for REBELS-04, the continuum detection increased from $3.4\sigma$ \citep{Inami2022} to $3.8\sigma$, providing further evidence that dust continuum emission from this source is real.

To estimate the IR luminosities, we assume that the dust SED can be represented by a simple modified blackbody (MBB) with a $T_{\text{dust}} = 47$ K and $\mathrm{\beta_{IR}} = 2.03$, analogous to the work of \citet{sommovigo2022}. \citet{sommovigo2022} find a scatter of 6 K on the median dust temperature of the REBELS sources and we combine this scatter in quadrature with the uncertainty on the IR luminosity. We correct for the effect of the CMB on the measured continuum flux according to the prescription of \citet{dacunha2013} assuming the dust is in thermal equilibrium.  Assuming a dust temperature of 47~K, we consider both the CMB heating the dust as well as the fact that the dust continuum is measured against the CMB. The IR luminosity is calculated by integrating the dust SED from 8 to 1000 $\mu$m.

Additionally, we calculate the IR luminosity with a higher dust temperature of $T_{\text{dust}} = 70$ K, to understand the effect of potentially high dust temperatures. This is motivated by $z>8$ galaxies with evidence of dust temperatures in excess of 70 K (e.g. \citealt{bakx2020}) as well as by the previous work on the REBELS galaxies of \citet{Inami2022}. A MBB with a dust temperature of approximately 70 K results in equivalent IR luminosities to the treatment adopted by \citet{Inami2022} for the [OIII]$_{88\mu\text{m}}$ targets. If the dust temperatures of the [OIII]$_{88\mu\text{m}}$ sources are $\gtrsim$ 70 K, the IR luminosities in this work would increase by a factor of $\gtrsim$ 3.

\begin{table*}
    \centering
    \begin{tabular}{cccccccc}
    \hline
    REBELS ID &  $z_{\mathrm{phot}}^{a}$ & $z_{\mathrm{[CII]}}^{b}$ &  $\mathrm{S_{cont,\ Band\ 6}}$   & $\mathrm{S_{cont,\ Band\ 7}}$  & $F_{\mathrm{[OIII]_{88\mu\text{m}},\ 100\ km\ s^{-1}}}  $ & $F_{\mathrm{[OIII]_{88\mu\text{m}},\ 400\ km\ s^{-1}}}  $ \\
     & &  & ($\mu$Jy) & ($\mu$Jy) & $(\mathrm{Jy\  km\ s^{-1})}$ & $(\mathrm{Jy\  km\ s^{-1})}$ \\ \hline
    REBELS-04  &  $8.57^{+0.10}_{-0.09}$ & - & - & $72 \pm 19$ & $<0.148$ & $<0.272$ \\[0.8ex] 
    REBELS-11  & $8.24^{+0.65}_{-0.37}$ & - & -& <97 & $<0.136$ & $<0.254$ \\[0.8ex]
    REBELS-13 &  $8.19^{+0.84}_{-0.50}$ & - & - & <88 & $<0.252$ & $<0.467$\\[0.8ex]
    REBELS-37  &  $7.75^{+0.09}_{-0.17}$ & $7.6428 \pm 0.0005$ & $<50$ & $100 \pm 17$ & - & - \\[0.8ex]
    \hline
    \end{tabular}
    \caption{The REBELS sources that have been scanned for [OIII]$_{88\mu\text{m}}$ emission and their corresponding continuum and [OIII]$_{88\mu\text{m}}$ fluxes$^{*}$. We provide upper limits ($5\sigma$) on the [OIII]$_{88\mu\text{m}}$ flux by assuming that the [OIII]$_{88\mu\text{m}}$ frequency is in the SPWs observed with ALMA. Continuum fluxes in in this table are without correction for the CMB and in the case of a non-detection are $3\sigma$ upper limits. \\ 
    $^{a}$ From \citet{REBELS}. If we include the effect of the CGM on the redshift likelihood distribution, the redshift of REBELS-04 could be smaller: $z_{\mathrm{phot}} = 8.43^{+0.10}_{-0.10}$ (Appendix~\ref{sec:phot_redshift}).\\
    $^{b}$ from Schouws et al. in preparation; see also \citet{harikane2024}.\\
    $^{*}$ The [OIII]$_{88\mu\text{m}}$ fluxes are tentative as these sources do not have spectroscopic confirmation of the photometric redshift yet. Note that the upper limits on the [OIII]$_{88\mu\text{m}}$ fluxes of REBELS-11 and REBELS-13 are particularly preliminary as only $\sim$40-50 $\%$ of the redshift likelihood distribution is covered by the SPWs.}
    \label{tab:sources}
\end{table*}

\begin{table*}
    \centering
    \begin{tabular}{ccccccc}
    \hline
    REBELS ID  & $\mathrm{SFR_{UV}}^{a}$ & $L\mathrm{_{IR}}$ & $\mathrm{SFR_{IR}}$ &  $\mathrm{SFR_{UV+IR}}$ & $L\mathrm{_{[OIII]_{88\mu\text{m}}}}$& $\mathrm{SFR_{[OIII]_{88\mu\text{m}}}}$\\
      &  ($\mathrm{M_{\odot}\ yr^{-1}}$)& ($10^{11} \mathrm{L_{\odot}}$) & ($\mathrm{M_{\odot}\ yr^{-1}}$)  & ($\mathrm{M_{\odot}\ yr^{-1}}$)& ($10^8  \mathrm{L_{\odot}}$) & ($\mathrm{M_{\odot}\ yr^{-1}}$)  \\ \hline
    REBELS-04 &   $23 \pm 2$ & $1.4_{-0.5}^{+0.6}$ & $17^{+8}_{-6}$& $40^{+8}_{-6}$ &$<4.2\ (<8.1)$& $<18\ (<36)$\\
    REBELS-11 &   $39 \pm 8$ & $< 1.9$  & $< 23$ & $<62$ &$<3.9\ (<7.6)$ & $<17\ (<33)$   \\
    REBELS-13 &   $44\pm 9$ & $< 1.7$ & $<21$ & $<65$ & $<7.4\ (<14.1)$& $<33\ (<64)$  \\
    REBELS-37  & $28 \pm 3$ &  $2.0_{-0.7}^{+0.9}$ & $24^{+11}_{-8}$ & $52^{+11}_{-8}$ & - & -  \\
    \hline
    \end{tabular}
    \caption{The IR, [OIII]$_{88\mu\text{m}}$ and UV luminosities and SFRs of the four REBELS sources of this work. We show the $5 \sigma$ upper limits on [OIII]$_{88\mu\text{m}}$ luminosities and SFR for different assumed FWHMs of the line: 100 km s$^{-1}$ (400 km s$^{-1}$). These limits are reliant on the redshifts of these targets lying within the observed scan range for [OIII]$_{88\mu\text{m}}$, which is less certain for REBELS-11 and REBELS-13. The SFR calculated from [OIII]$_{88\mu\text{m}}$ is based on the relation from $z=6$-9 galaxies derived by \citet{harikane2020}. We note that a decrease in the photometric redshift of REBELS-04 due to an additional CGM component to $z=8.43$ would decrease the IR luminosity by $\approx$3$\%$ only. IR luminosities are estimated assuming a modified blackbody with a dust temperature of 47 K. If a dust temperature of $\gtrsim70$ K is assumed, the IR luminosity is $\gtrsim3$ times higher.\\
    $^{a}$ From \citet{REBELS}.}
    \label{tab:luminosities}
\end{table*}

\section{Results}
\label{sec:results}
\subsection{Upper limits on $L_{\text{[OIII]$_{88\mu\text{m}}$}}$ of REBELS sources}
\label{sec:results1}
\begin{figure*}
    \includegraphics[width = \textwidth]{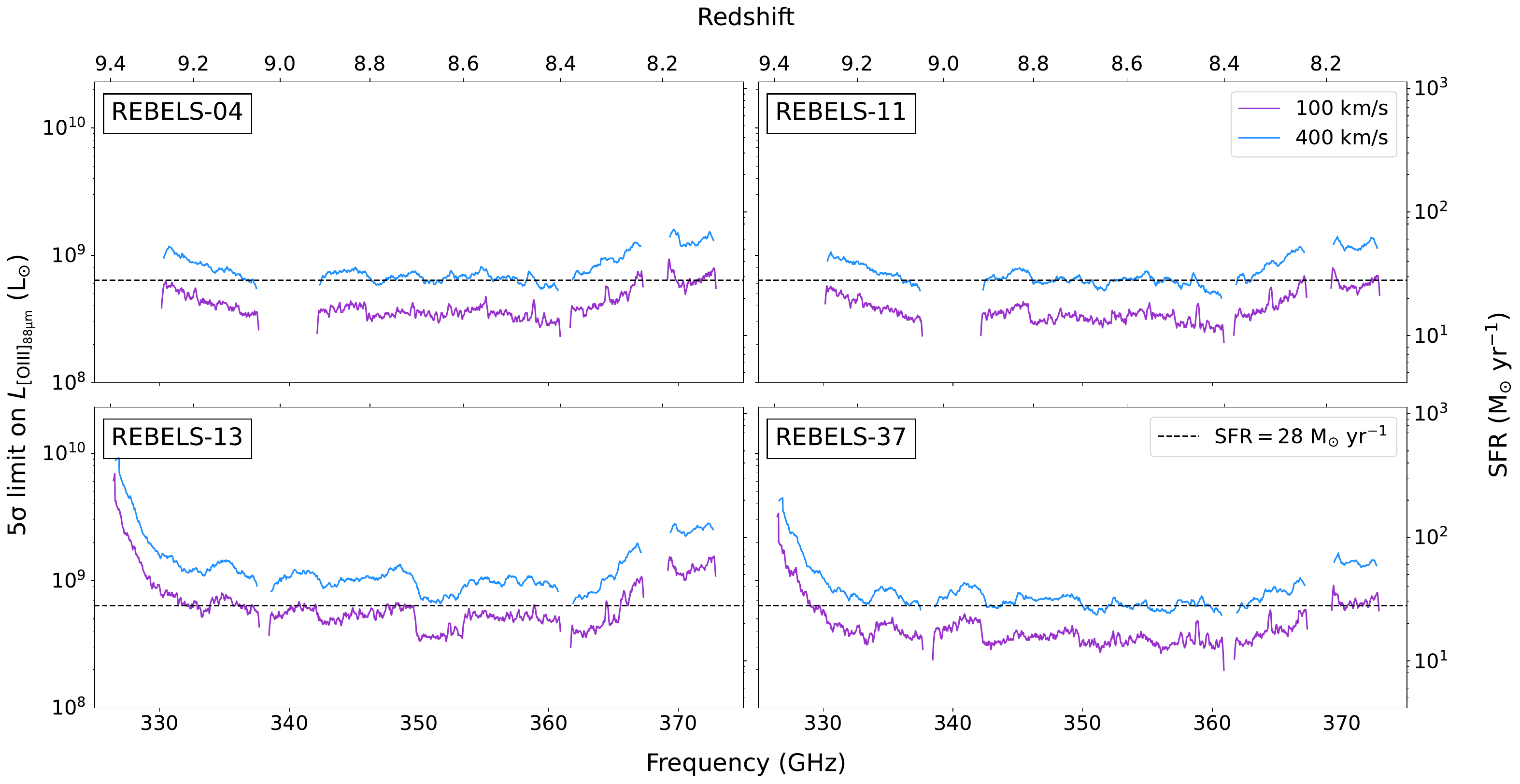}
    \caption{The 5$\sigma$ limit on the [OIII]$_{88\mu\text{m}}$ luminosity as function of frequency and redshift for the observations on the four REBELS targets (assuming a FWHM of 100 and 400 km s$^{-1}$). The horizontal dotted lines indicate the SFR limit of $28\ \text{M}_{\odot}\ \text{yr}^{-1}$, for which the REBELS survey has been most efficient in the detection of [CII]$_{158\mu\text{m}}$. [OIII]$_{88\mu\text{m}}$ luminosity is converted to SFR using the relation from \citet{harikane2020} derived from $z=6$-9 galaxies.}
    \label{fig:loiii}
\end{figure*}

The REBELS survey targeted [OIII]$_{88\mu\text{m}}$ line emission in four $z \gtrsim$~7.6 galaxies and, as indicated in Section \ref{sec:method_oiii}, we do not identify any line emission in the observations. To demonstrate the depth of our observations, we show the $5\sigma$ limit on the [OIII]$_{88\mu\text{m}}$ luminosity as a function of observed frequency in Fig.~\ref{fig:loiii}. To estimate the SFR limits these luminosities correspond to, we use the relation derived from $z=6$-9 galaxies by \citet{harikane2020}.\footnote{At the $5\sigma$ $L_{\text{[OIII]$_{88\mu\text{m}}$}}$ limit of our sample the SFRs that one obtains with the metal-poor local dwarfs relation from \citet{looze2014} will be a factor 0.8 smaller on average. If we assume the local starburst relation from \citet{looze2014} the resulting SFRs would be a factor 5.6 (4.5) larger on average assuming a FWHM of 100 km s$^{-1}$ (400 km s$^{-1}$). } For a FWHM of 100 and 400 km s$^{-1}$, the observations reach down to 3.9 $\times$ 10$^{8}$ $\text{L}_{\odot}$ and 7.6 $\times$ 10$^{8}$ $\text{L}_{\odot}$, respectively, equivalent to SFR $\sim 17$ and $33\ \text{M}_{\odot}\ \text{yr}^{-1}$. \citet{REBELS} show that at SFRs of $\gtrsim 28\ \text{M}_{\odot}\ \text{yr}^{-1}$ the REBELS survey has a 79$\%$ efficiency in detecting [CII]$_{158\mu\text{m}}$. We note that we do not utilize the upper limit on $L_{\text{[OIII]$_{88\mu\text{m}}$}}$ for REBELS-37 for further analysis, as the expected frequency for the [OIII]$_{88\mu\text{m}}$ line is outside of the frequency scan range. The upper limits on $L_{\text{[OIII]$_{88\mu\text{m}}$}}$ for the remaining sources are obtained from taking the average over all observed frequencies and are quoted in Table~\ref{tab:luminosities}. 

Also shown in Table~\ref{tab:luminosities} are the unobscured and obscured SFRs based on the UV and IR luminosities. For these SFRs we adopt the same relations as presented in \citet{REBELS}: $\mathrm{SFR_{UV}} =$~$7.1\times $~$10^{-29} \frac{L_{\nu}}{\mathrm{erg\ s^{-1}\ Hz^{-1}}}$ and $\mathrm{SFR_{IR}} = 1.2 \times 10^{-10} \frac{L_{\text{IR}}}{\text{L}_{\odot}}$. The unobscured SFR conversion factor is based on SED fitting of the REBELS galaxies using a \citet{Chabrier2003} initial mass function (IMF) \citep{topping2022}. The obscured SFR conversion factor is adopted from \citet{kennicutt1998} and converted to a \citet{Chabrier2003} IMF. We note that we use these conversion factors when comparing to galaxies from literature as well. 

Of the three targets (REBELS-04, REBELS-11 and REBELS-13) observed with ALMA Band 7 to scan for the [OIII]$_{88\mu\text{m}}$ emission line, only REBELS-04 has observations covering the majority of its redshift likelihood distribution ($\geq 92\%$), with $z_{\mathrm{phot}} = 8.57^{+0.10}_{-0.09}$ ($z_{\mathrm{phot}} = 8.43^{+0.10}_{-0.10}$) without (with) an additional correction for absorption by the CGM.  We note that this estimate leverages both the existence of both an apparent Lyman break in the SED and a contribution of line emission to the 4.5$\mu$m flux (see Appendix~\ref{sec:phot_redshift}).

\begin{figure}
    \includegraphics[width = 0.5\textwidth]{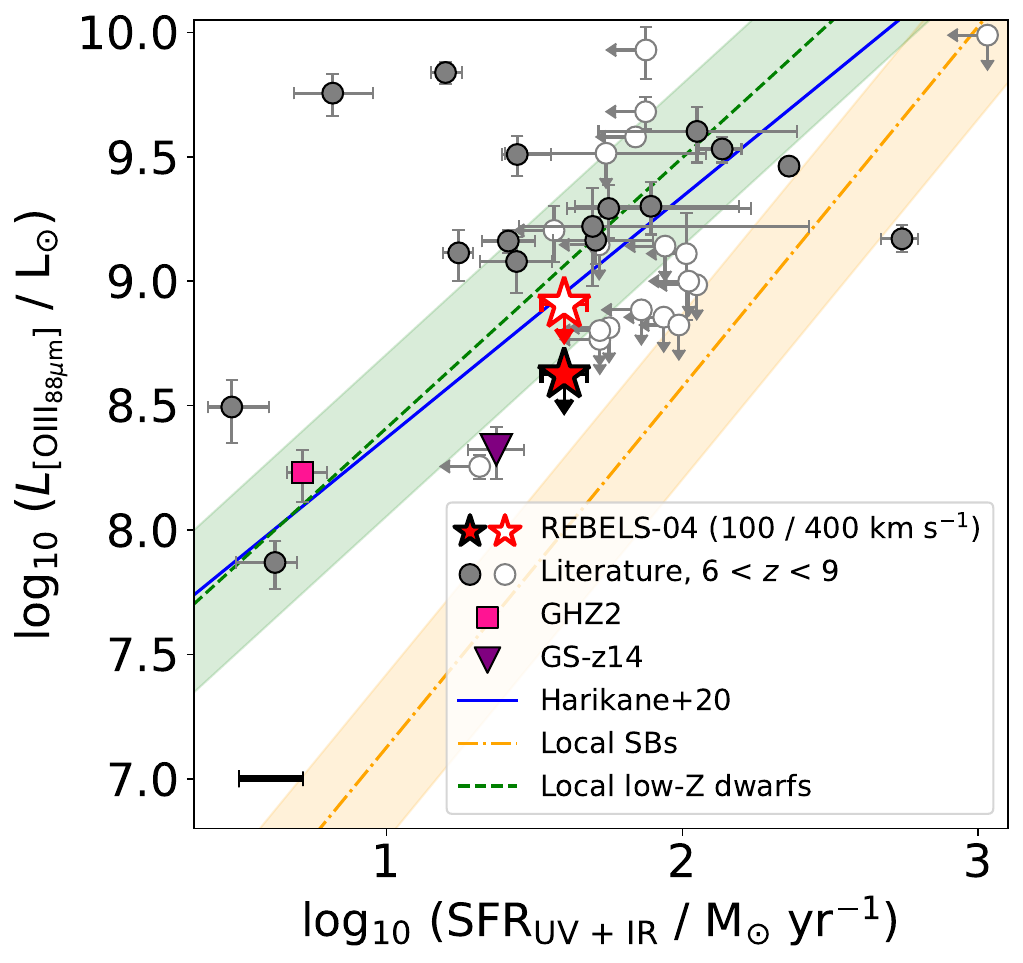}
    \caption{The [OIII]$_{88\mu\text{m}}$ luminosity as function of total SFR (UV+IR). The [OIII]$_{88\mu\text{m}}$ upper limits of REBELS-04 are shown by the stars, assuming either a FWHM of 100 km  s$^{-1}$ (filled) or 400 km  s$^{-1}$ (open).  The plotted constraints assume the photometric redshift estimate $z_{\mathrm{phot}} = 8.57^{+0.10}_{-0.09}$ for REBELS-04 is accurate and lying within the scan range of the REBELS [OIII]$_{88\mu\text{m}}$ observations. Literature observations of [OIII]$_{88\mu\text{m}}$ are plotted with circles and are filled in the case of a detection of both [OIII]$_{88\mu\text{m}}$ and dust continuum. The two recent detections of [OIII]$_{88\mu\text{m}}$ at $z > 12$ in GHZ2 and GS-z14 are shown with the square and triangle, respectively \citep{zavala2024, schouws2024, carniani2024}.  The approximate shift in the total SFR when assuming a dust temperature that is 10~K lower/higher than 47~K is shown by the thick errorbar in the lower left corner. The \mbox{$L_{\text{[OIII]$_{88\mu\text{m}}$}}$-SFR} relation derived by \citet{harikane2020} based on $z= 6$-9 galaxies and the relations from \citet{looze2014} derived with local starburst and low-metallicity dwarf galaxies are indicated with the solid and dashed lines.}
    \label{fig:L_SFR}
\end{figure}

Unfortunately, no line detection is found for any targets over the full scan range. While the observations reach down to similar SFRs to which [CII]$_{158\mu\text{m}}$ detections were commonly found in REBELS, it is possible this limit is not sufficient for [OIII]$_{88\mu\text{m}}$. This would imply targets have a lower $L_{\text{[OIII]$_{88\mu\text{m}}$}}$ than expected according to their SFR. In Figure \ref{fig:L_SFR} we plot the $5\sigma$ upper limits on $L_{\text{[OIII]$_{88\mu\text{m}}$}}$ obtained with both FWHMs (100 and 400 km s$^{-1}$) for REBELS-04. Interestingly we see that when we adopt a FWHM of 100 km s$^{-1}$ the upper limit on $L_{\text{[OIII]$_{88\mu\text{m}}$}}$ of REBELS-04 is significantly below the luminosities found for [OIII]$_{88\mu\text{m}}$-detected literature galaxies at $z>6$ with similar SFRs.

Also plotted are the $z = 6$-9 galaxies from the literature \citep{inoue2016, Carniani2017, hashimoto2018, marrone2018, walter2018, hashimoto2019, tamura2019, harikane2020, akins2022, tadaki2022, witstok22, wong2022, ren2023, algera_oiii_cii, bakx2024, carniani2024, fujimoto2024, schouws2024, zavala2024}. For these literature galaxies we recalculated the IR luminosity based on the dust continuum at $88\ \mu \text{m}$ (corrected for the CMB) if there is only one band continuum observation available with a modified blackbody with $T_{\text{dust}} = 47$~K and $\beta_{\text{IR}} = 2.03$. If a source has an upper limit on the IR SFR, we adopt the total SFR from the literature derived by SED fitting if those constraints are not limits. The approximate difference in the total SFR when assuming a dust temperature that is 10 K higher or lower than 47 K is indicated with the errorbar in the lower left corner in Figure~\ref{fig:L_SFR}.

\subsection{[OIII]$_{88\mu\text{m}}$ deficit}
In Section \ref{sec:method_dust} we discuss that we observe significant dust continuum emission in Band 7 for REBELS-04 and REBELS-37. These dust detections are  $\sim 3.8 \sigma$ and $\sim 6 \sigma$, respectively. The resulting IR luminosities and upper limits of the non-detections are reported in Table~ \ref{tab:luminosities}. Local dwarf and star-forming galaxies are known to exhibit an anti-correlation between the [OIII]$_{88\mu\text{m}}$/IR luminosity ratio and IR luminosity, also known as the [OIII]$_{88\mu\text{m}}$ deficit (e.g. \citealt{madden2013, Cormier2015}). Galaxies and QSOs at $z > 7$ have been found to show an analogous [OIII]$_{88\mu\text{m}}$ deficit \citep{tamura2019, hashmimoto2019}. 

\begin{figure}
    \includegraphics[width = 0.5\textwidth]{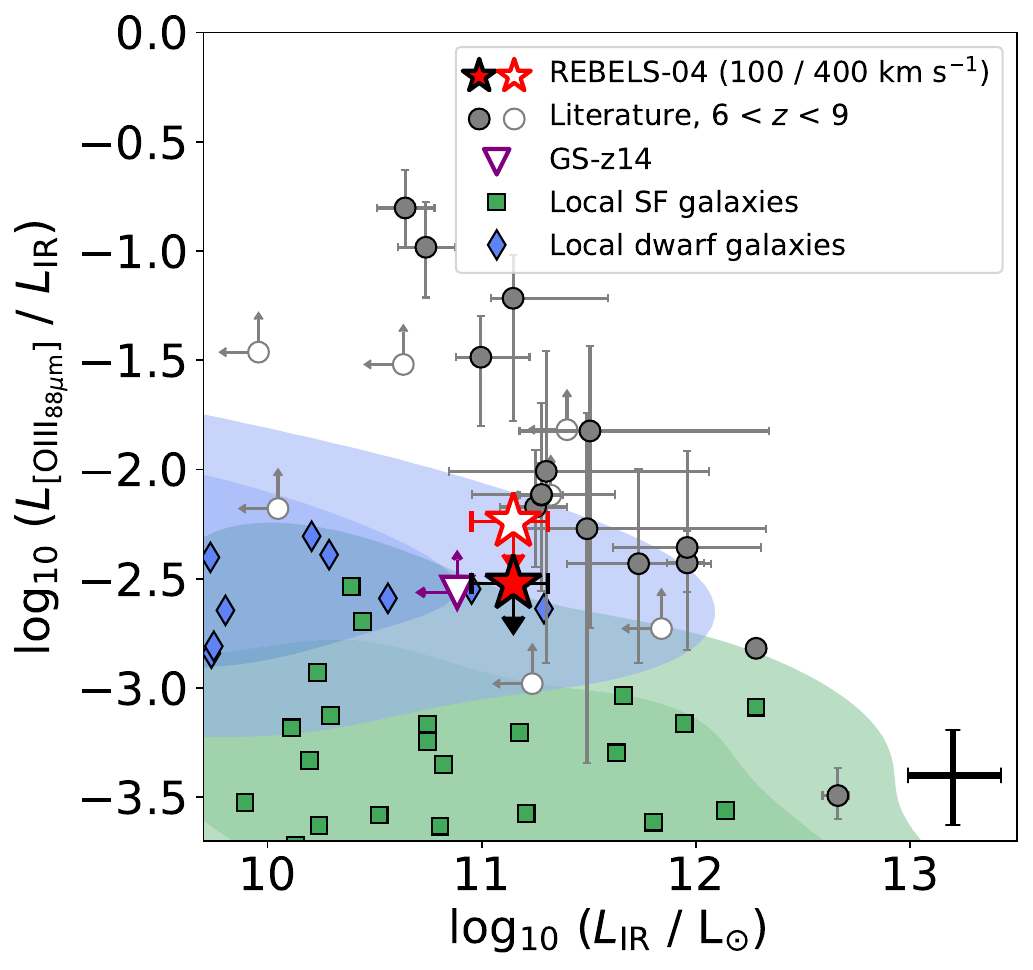}
    \caption{The [OIII]$_{88\mu\text{m}}$ to IR luminosity ratio as a function of IR luminosity. The only source with an upper limit on [OIII]$_{88\mu\text{m}}$ and a dust detection, REBELS-04, is plotted with the star for two different assumed FWHMs. Galaxies at $z = 6-9$ from the literature are plotted with circles (filled when detected in both [OIII]$_{88\mu\text{m}}$ and dust continuum) and their IR luminosities are recalculated with a modified blackbody with $T\mathrm{_{dust}} = 47$ K if there is a dust continuum observation in only one ALMA band. The errorbar in the lower right corner shows the change on the data points if they have a dust temperature that is 10 K lower of higher than 47 K. The recent detection of [OIII]$_{88\mu\text{m}}$ at $z\sim 14$ is shown with a triangle \citep{schouws2024, carniani2024}. Local star forming \citep{herrara2018, herrara2018_2} and dwarf galaxies \citep{madden2013, Cormier2015} are plotted by the squares and diamonds, respectively. The contours show the 68$^{\text{th}}$ and 95$^{\text{th}}$ percentiles of both distributions of local galaxies.}
    \label{fig:OIII_IR}
\end{figure}

As we have a continuum detection for REBELS-04, we can explore the [OIII]$_{88\mu\text{m}}$ deficit of this REBELS source. In Figure \ref{fig:OIII_IR} the luminosity ratio of REBELS-04 is shown for the two assumptions on the FWHM. Also plotted are the same $z = 6$-9 galaxies from the literature as shown in Figure \ref{fig:L_SFR}. Moreover, we show local star-forming galaxies from SHINING \citep{herrara2018, herrara2018_2} and local dwarf galaxies \citep{madden2013, Cormier2015} adopted from Figure 4 from \citet{tamura2019} and the 68$^{\text{th}}$ and 95$^{\text{th}}$ percentiles of both distributions. The [OIII]$_{88\mu\text{m}}$ deficit of REBELS-04 when assuming a FWHM of 100 km s$^{-1}$ is potentially more consistent with local star-forming galaxies than sources at $z = 6$-9 with similar IR luminosities. We note that if the dust temperature for REBELS-04 were higher than 47 K, it would lead to a higher IR luminosity and would make the discrepancy between REBELS-04 and other $z = 6$-9 galaxies larger.

\subsection{ALMA Band 7 continuum detections}
\begin{figure*}
    \includegraphics[width = \textwidth]{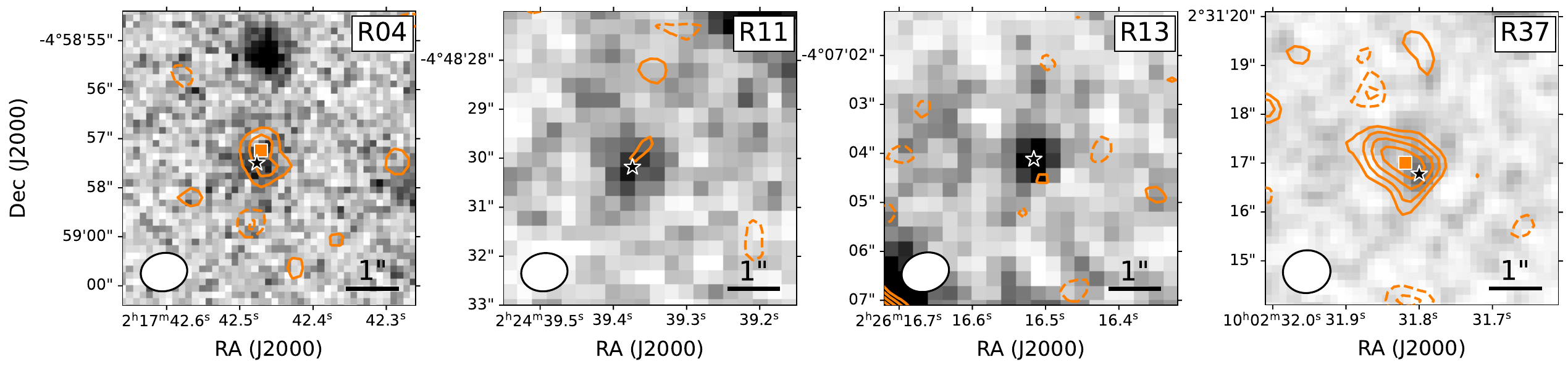}
    \caption{Observations of the dust continuum at $\sim 88 \mu$m for the four galaxies in REBELS targeted with [OIII]$_{88\mu \text{m}}$ line scans with ALMA. The dust continuum is shown with the contours from 2 to $5\sigma$ (dashed contours are negative), the background image is the astrometrically corrected rest-UV (JHK-stack) images from \citet{Inami2022}. The squares and stars show the peaks of the dust continuum and rest-UV emission, respectively. The ALMA beam is shown in the left, lower corners of the figures.  While \citet{Inami2022} already presented the full depth continuum detection of REBELS-37, additional Band 7 data became available for REBELS-04 and REBELS-11 in Cycle 8, increasing the significance of the continuum detection of REBELS-04 to $3.8\sigma$. REBELS-04 and REBELS-37 represent two of the highest redshift galaxies showing significant dust detections.}
    \label{fig:continuum}
\end{figure*}

The Band 7 dust continuum detection of two $z > 7.6$ galaxies is noteworthy. If the redshift of $z\sim 8.6$ of REBELS-04 can be confirmed, it would be the highest redshift ALMA detection of the dust continuum (c.f. MACS0416$\_$Y1 at $z$ = 8.31; \citealt{tamura2019, bakx2020, Tamura2023}). In Figure \ref{fig:continuum} we show the dust continuum contours (2 to 5$\sigma$) of the REBELS sources on top of the astrometry corrected NIR (rest-UV) images from \citet{Inami2022}. We clearly see no significant dust-emission for REBELS-11 and REBELS-13 at their rest-UV locations, only a small $2\sigma$ signal. Analogous to \citet{Inami2022} we have fitted the continuum and rest-UV emission with a 2D Gaussian to determine the location of the peak of the emission. The peaks of the dust continuum and rest-UV emission are indicated with the squares and stars, respectively. Moreover, the ALMA beam is plotted in the left lower corners.

In high redshift galaxies, it is relatively common that UV-bright regions and the regions emitting dust continuum emission are not cospatial, therefore allowing the galaxy to be UV-bright but still host a significant dust reservoir (e.g. \citealt{Carniani2017, Bowler2022, hygate2023, rowland2024}). If the rest-UV and dust continuum regions are offset this can for example imply physically different regions of the galaxy, or a minor/major merging (e.g. \citealt{behrens2018, pallottini2022}). At the resolution of the current ALMA observations we cannot determine if such an offset is present in the targets of this work.

\subsection{Constraints on $T_{\text{dust}}$ using Band 6 and 7}
\label{sec:tdust}
To correctly interpret the FIR emission and convert this to an obscured star formation rate, the characterization of the temperature of the dust ($T_{\text{dust}}$) and the emissivity index ($\beta_{\text{IR}}$) are crucial.  While it is hard to constrain $T_{\text{dust}}$ and $\beta_{\text{IR}}$ from dust continuum observations in only one ALMA band (see e.g. \citealt{inoue2020, sommovigo2022, fudamoto2023} for methods that estimate $T_{\text{dust}}$ when assuming a $\beta_{\text{IR}}$), when combining the observations in multiple ALMA bands we can get constraints on the dust mass (e.g. \citealt{algera2024}). The dust temperature is thought to increase with redshift (e.g. \citealt{schreiber2018, liang2019,  bouwens2020, sommovigo2022, Mitsuhashi2024}). However, a large scatter of dust temperatures is found at $z > 7$, from relatively cold dust ($\sim 30$ K: \citealt{witstok22,algera2024}) to much hotter dust temperatures ($> 80$ K: \citealt{bakx2020}). 

For one source, REBELS-37, we have both Band 6 and 7 continuum observations and we therefore examine the dust temperature. The Band 7 observation of REBELS-37 show a $\sim 6 \sigma$ continuum detection, whereas in Band 6 we have an upper limit on the continuum flux. A detection at shorter wavelengths and a non-detection at longer wavelengths could be the result of a source with relatively hot dust. To investigate this the common approach used is to fit a modified blackbody SED to the continuum fluxes. We follow the method of \citet{algera_oiii_cii} but due to the relatively shallow limit in Band 6 we cannot obtain tight constraints on $T_{\text{dust}}$. The obtained $T_{\text{dust}} = 114^{+51}_{-52}$ K is highly dependent on the assumed prior on  $T_{\text{dust}}$. We discuss this procedure further in Appendix ~\ref{sec:sed_fit}. 

\begin{figure*}
    \centering
    \includegraphics[width = 0.95 \textwidth]{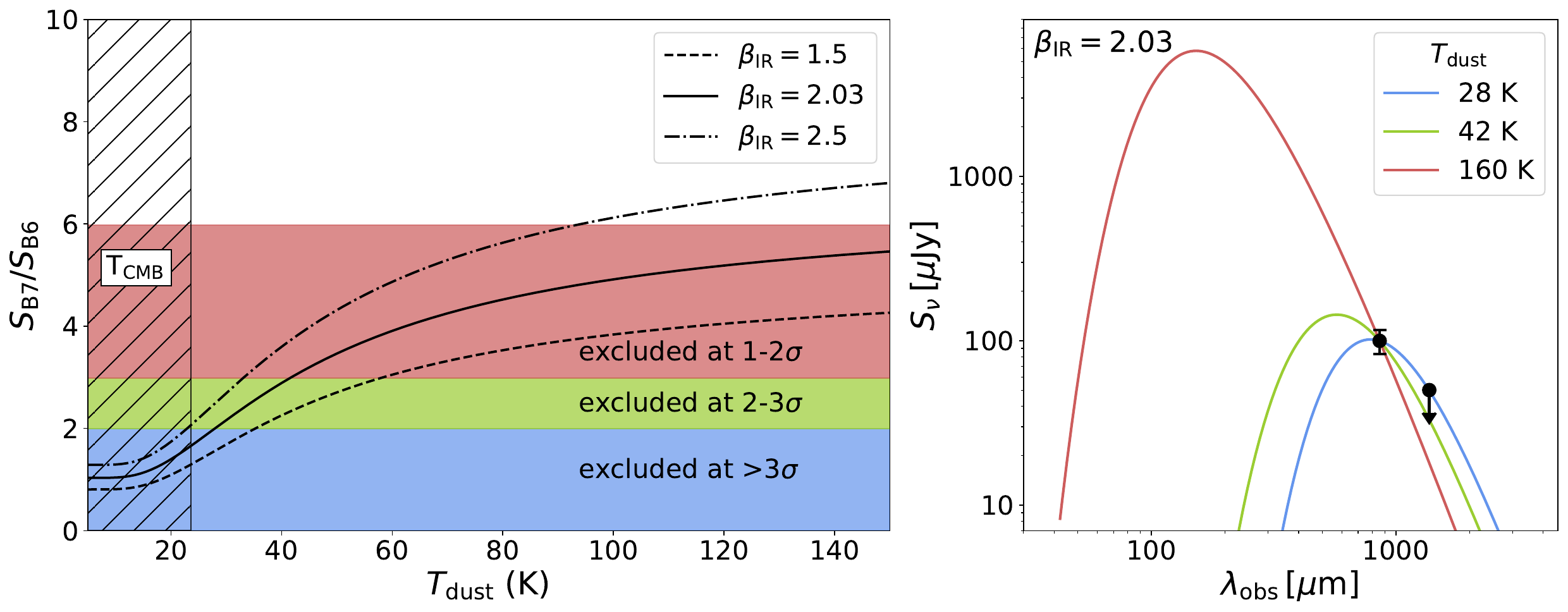}
    \caption{\textit{Left:} The ratio of the Band 7 and Band 6 continuum flux of REBELS-37 as a function of dust temperature plotted for distinct values of the emissivity index ($\beta_{\text{IR}})$. The Band 6 upper limit and Band 7 continuum detection of REBELS-37 exclude part of the parameter space, depending on the choice of upper limit on the Band 6 continuum flux, be it 1, 2, or 3$\sigma$. The underlying CMB temperature of $\approx 24$ K at $z \sim 7.6$ hinder detections of dust temperatures at lower temperatures. \textit{Right:} The dust SED plotted assuming $\beta_{\text{IR}} = 2.03$ for the dust temperatures limits that are identified with the Band 7 and Band 6 continuum flux ratio at 1, 2, and $3\sigma$ (red, green, and blue lines, respectively). The size of the errorbar on the upper limit is $1\sigma$. We can exclude a cold dust temperature of $\lesssim 28$~K at $3\sigma$ confidence.}
    \label{fig:ratio_sed}
\end{figure*}

Alternatively, we investigate the ratio of the continuum flux in Band 7 and 6 that is sensitive to only $T_{\text{dust}}$ and $\beta_{\text{IR}}$. In Figure \ref{fig:ratio_sed} we show the dependence of the continuum flux ratio on the dust temperatures for an assumed emissivity index. Also shown is the parameter space allowed by 1, 2, and 3 $\sigma$ upper limits on the Band 6 continuum flux. Assuming $\beta_{\text{IR}} = 2.03$ we can obtain an lower limit on the dust temperature. At $3 \sigma$ we can exclude $T_{\text{dust}} \lesssim  28$ K for REBELS-37, also shown by the dust SED plotted in the right panel of Figure \ref{fig:ratio_sed}. Dust temperatures lower than 42 K can be excluded at $2 \sigma$ confidence.\footnote{At only $1 \sigma$ certainty, dust temperatures below 160 K are excluded.} From these results we can infer that it is likely that the REBELS-37 is not as cold as REBELS-25, which has a dust temperature of 32 K \citep{algera2024}. 

\section{Discussion}
\label{sec:discussion}

\subsection{[OIII]$_{88\mu\text{m}}$ non-detections}
\subsubsection{REBELS-04}
All four REBELS sources that were observed with ALMA in Band 7 do not show [OIII]$_{88\mu\text{m}}$ line emission. The lack of an [OIII]$_{88\mu\text{m}}$ detection in REBELS-04 is especially interesting. The redshift likelihood distribution of REBELS-04 is covered by $\geq 92\%$ with the ALMA observations and the constraints on the redshift (\mbox{$z_{\text{phot}} = 8.57^{+0.10}_{-0.09}\pm0.10$}) are tight due to additional \textit{HST} observations. As shown in Figure \ref{fig:loiii}, our observations reach $5\sigma$ depths of $< 18\ \text{M}_{\odot}\ \text{yr}^{-1}$ ($< 36\ \text{M}_{\odot}\ \text{yr}^{-1}$), according to the SFR relation from \citet{harikane2020} and assuming a FWHM of $100$~km~s$^{-1}$ ($400$~km~s$^{-1}$). Considering that the total SFR (UV + IR) of \mbox{REBELS-04} is $40^{+8}_{-6}\ \text{M}_{\odot}\ \text{yr}^{-1}$ we would expect the depth of the observations to be sufficient under the assumption of a FWHM of $100$ km s$^{-1}$. However, as we observe in Figures \ref{fig:L_SFR} and \ref{fig:OIII_IR} the upper limit on the [OIII]$_{88\mu\text{m}}$ luminosity is generally lower than the luminosities from other sources at high redshift. It is worth noting that if the dust temperature is in excess of 47 K, this discrepancy would be even larger due to the higher SFR and IR luminosity implied for REBELS-04.

The requested depth of the REBELS survey is based on previous REBELS pilot programs targeting [CII]$_{158\mu\text{m}}$ emission in EoR galaxies \citep{smit2018, schouws2023}. To determine the sensitivity requirement for [OIII]$_{88\mu\text{m}}$, an [OIII]$_{88\mu\text{m}}$/[CII]$_{158\mu\text{m}}$ luminosity ratio of 3.5 is assumed (see \citealt{REBELS}). Interestingly, \citet{algera_oiii_cii} finds luminosity ratios \mbox{[OIII]$_{88\mu\text{m}}$/[CII]$_{158\mu\text{m}}$ $\approx 1$-1.5} for two REBELS galaxies with multiple ALMA band observations that are more in line with predictions from high-$z$ simulations (e.g. \citealt{katz2022, pallottini2022, schimek2024}). If the [OIII]$_{88\mu\text{m}}$/[CII]$_{158\mu\text{m}}$ ratio is indeed $\approx 1$-1.5 (for example due to a higher photodissociation region (PDR) covering fraction, see e.g.  \citealt{harikane2020}), the [OIII]$_{88\mu\text{m}}$ luminosity based of the [CII]$_{158\mu\text{m}}$ luminosity will decrease with a factor of 2-3. That could make the current sensitivity of the observations too shallow to detect the faint [OIII]$_{88\mu\text{m}}$ emission. For REBELS-04, which has good constraints from its photometric redshift, it seems very likely this is the case.

To understand which ISM conditions can lead to faint [OIII]$_{88\mu\text{m}}$ emission other FIR emission lines are needed. For the targets in this work unfortunately no other FIR emission lines have been observed with ALMA. However, we will briefly mention ISM conditions that could lead to a decrease in [OIII]$_{88\mu\text{m}}$ luminosity. A low metallicity can explain the faintness of [OIII]$_{88\mu\text{m}}$ due to less oxygen that is available to be ionized. \citet{jones2020} have calibrated a relation between the metallicity, [OIII]$_{88\mu\text{m}}$ luminosity and SFR of galaxies at $z\sim 8$. The \citet{jones2020} calibration adopts fiducial values of $T_{e} = 1.5 \times 10^4$ K and $n_{e} = 250$ cm$^{-3}$. For REBELS-04 we have good constraints on the total SFR from both UV and IR detections and find $\text{log}_{10} (L_{\text{[OIII]$_{88\mu\text{m}}$}}/\text{SFR}) < 7.0\  (7.3) $ assuming a FWHM of 100 km s$^{-1}$ (400 km s$^{-1}$). This results in an upper limit on $12+\text{log} (\text{O}^{++}/\text{H}^{+}) < 7.8\ (8.0) \approx  0.12\ (0.23)\ Z_{\odot}$ for REBELS-04 for a FWHM of 100 km s$^{-1}$ (400 km s$^{-1}$).\footnote{Here we use a solar oxygen abundance of $ 12+\text{log} (\text{O}^{++}/\text{H}^{+}) = 8.69$ \citep{asplund2009}.} We note that this estimated metallicity is lower than what has been found for other REBELS galaxies observed with \textit{JWST}, that have an average metallicity of $0.4\ Z_{\odot}$ \citep{rowland2025}.  It is challenging to reconcile the low metallicity of REBELS-04 with the detection of dust in the source, unless most of the metals are bound up in the dust and are not in the gas phase. Evidence from other REBELS galaxies that have been covered by \textit{JWST} observations show that on average only 16$\%$ of the metals are locked up in dust \citep{algera2025}, it is therefore likely that this cannot be the explanation for the faint [OIII]$_{88\mu\text{m}}$ luminosity of REBELS-04.

We note that the \textsc{cloudy} models from \citet{witstok22} show that the [OIII]$_{88\mu\text{m}}$ luminosity is very sensitive to age of the stellar population (decreasing by 0.5 dex for a burst age increase from 1 to 5~Myr).  This occurs due to the short lifetimes of the massive stars that produce the high-energy radiation to doubly ionize oxygen.  Along the same lines, the models of \citet{witstok22} indicate higher \mbox{[OIII] + H$\beta$} equivalent widths (EW([OIII] + H$\beta$)) for younger starbursts. The EW([OIII] + H$\beta$) of REBELS-04 is relatively high at $\sim$$1800\ \text{\AA} $ \citep{bouwens2020}, pointing to a younger stellar population age for the source.  The faintness of the [OIII]$_{88\mu\text{m}}$ emission from REBELS-04 is therefore unlikely the result of the age of the starburst.

As discussed in \citet{harikane2020}, the ISM density, ionization parameter and metallicity all have an impact on the [OIII]$_{88\mu\text{m}}$ luminosity (and the [OIII]$_{88\mu\text{m}}$/[CII]$_{158\mu\text{m}}$ luminosity ratio). The \textsc{cloudy} models of \citet{harikane2020} use a 1 Myr old bursty star formation model and a \citet{salpeter1955} IMF (1-100 M$_{\odot}$). When we compare the $L_{\text{[OIII]$_{88\mu\text{m}}$}}/\text{SFR}$ ratio of REBELS-04 to the \textsc{cloudy} models of \citet{harikane2020} there is no clear solution to the ISM conditions ($\text{U}_{\text{ion}}$, $n_{\text{H}}$ and $Z$) as we are working with upper limits.  However, we can estimate what the change in these ISM parameters should be for REBELS-04 to be consistent with the [OIII]$_{88\mu\text{m}}$ luminosities from other $z=6$-9 sources, according to the \citet{harikane2020} $L_{\text{[OIII]$_{88\mu\text{m}}$}}$-SFR relation. Adopting a FWHM of 100 and 400~km~s$^{-1}$, the [OIII]$_{88\mu\text{m}}$ luminosity of REBELS-04 is at least 0.33 and 0.04 dex lower than the [OIII]$_{88\mu\text{m}}$ luminosity implied by the total SFR. A decrease in the [OIII]$_{88\mu\text{m}}$ luminosity of at least 0.5~dex can be achieved by a decrease in metallicity from 0.2 to 0.05~$Z_{\odot}$ according to the \textsc{cloudy} models of \citet{harikane2020}. As we discussed above, we do not expect the metallicity to be the main reason for the faint [OIII]$_{88\mu\text{m}}$ emission as REBELS-04 has a dust continuum detection.

\begin{figure}
    \centering
    \includegraphics[width = 0.5 \textwidth]{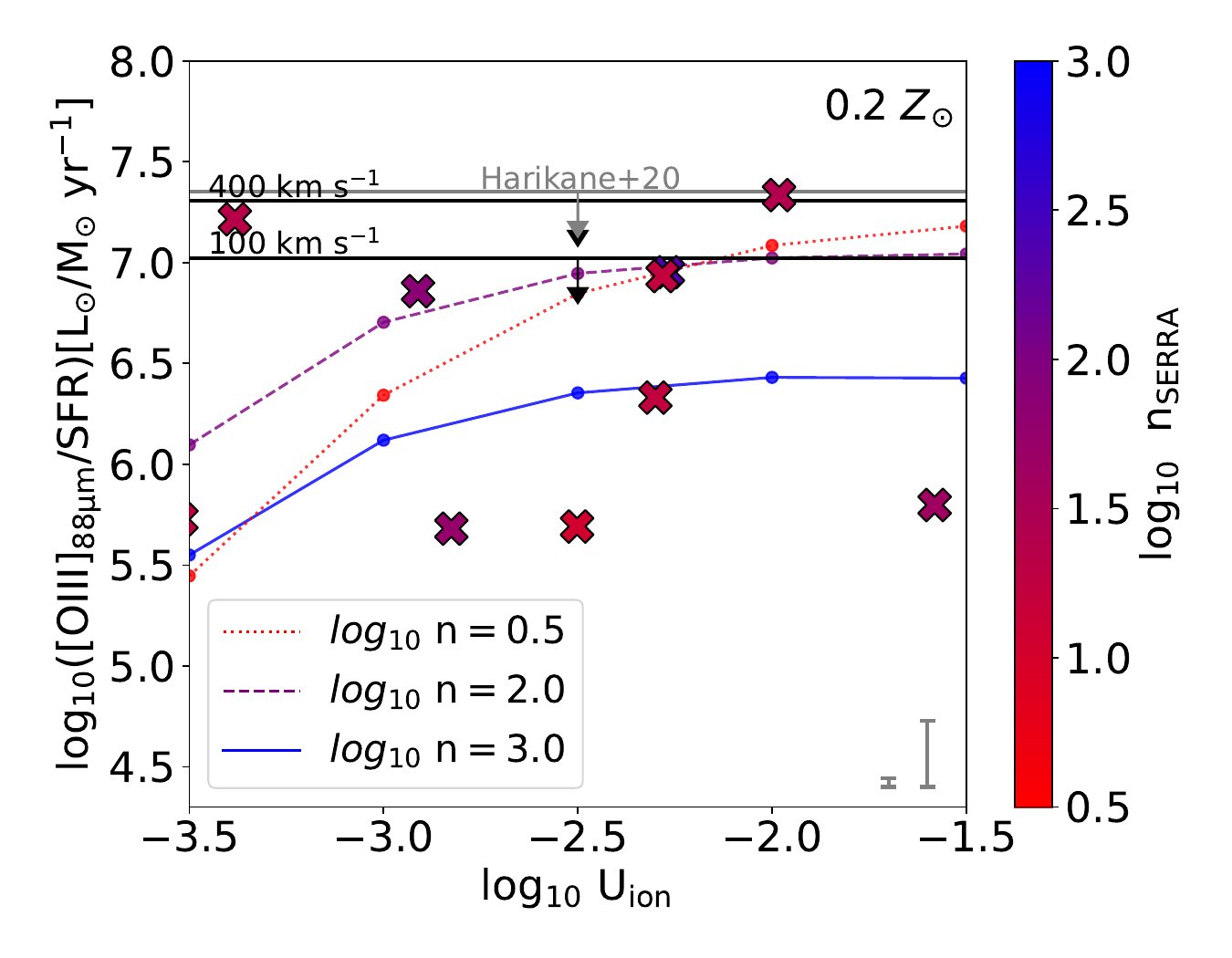}
    \caption{The [OIII]$_{88\mu\text{m}}$ luminosity to SFR ratio as function of the ionization parameter. The upper limits of REBELS-04 for a FWHM of 100 and 400 km s$^{-1}$ are shown as well as the upper limit the source should have to lie on the \citet{harikane2020} $\mathrm{L_{[OIII]_{88\mu\text{m}}}}$-SFR relation. The errorbars in the lower right corner show the deviation from the \citet{harikane2020} relation for both FWHMs. The \textsc{cloudy} models for $0.2\ Z_{\odot}$ of \citet{harikane2020} are plotted with the differently colored solid and dotted lines for different densities. Additionally, the SERRA galaxies at $z \sim 7.7$ are shown with the crosses color-coded by their ISM density \citep{pallottini2022}.}
    \label{fig:ism_parameters}
\end{figure}

In Figure~\ref{fig:ism_parameters} we show the $0.2\ Z_{\odot}$ \textsc{cloudy} models of \citet{harikane2020} (converted to a \citet{Chabrier2003} IMF) as function of the density and ionization parameter, as well as the parameter space that can be ruled out by our upper limits on the [OIII]$_{88\mu\text{m}}$ luminosity and the upper limit that REBELS-04 should have according to its SFR. A decrease of $\sim$0.5 dex in [OIII]$_{88\mu\text{m}}$ luminosity can also be achieved by increasing the density of hydrogen from 100 to 1000 cm$^{-3}$. This is due to collisional de-excitation that becomes important once the critical density of 510 cm$^{-3}$ is reached.\footnote{The average electron density is expected to increase with redshift and at $z \sim 8.6$ the electron density might be equal or even higher than the critical density of [OIII]$_{88\mu\text{m}}$ \citep{isobe2023}.} Therefore, a high density could cause the faint [OIII]$_{88\mu\text{m}}$ luminosity in REBELS-04.

From Figure~\ref{fig:ism_parameters} we can also see that the ionization parameter at $\mathrm{log_{10}\ U_{ion} \lesssim -2.5}$ can have a significant impact on the [OIII]$_{88\mu\text{m}}$ luminosity according to the \textsc{cloudy} models. Also plotted are the $z \sim 7.7 $ SERRA galaxies with $0.1\ Z_{\odot} < Z < 0.3\ Z_{\odot}$ from \citet{pallottini2022}. The [OIII]$_{88\mu\text{m}}$ luminosity of the SERRA galaxies does not show a clear dependence on the density in the same parameter space of the \textsc{cloudy} models. The [OIII]$_{88\mu\text{m}}$ luminosity of SERRA galaxies at $z = 11$- 14 seem to depend mainly on the ionization parameter \citep{Kohandel2023}. \citet{Kohandel2023} find that the SERRA galaxies that have the largest downward deviation from the local metal poor dwarf galaxies SFR relation \citep{looze2014} have $\mathrm{log_{10}\ U_{ion} < -3.0}$. A large dependence of [OIII]$_{88\mu\text{m}}$ on the ionization parameter has also been found in other high-$z$ simulations (e.g. \citealt{katz2022, nakazato2023}). Accordingly, the low [OIII]$_{88\mu\text{m}}$ luminosity of REBELS-04 might be caused by a low ionization parameter and/or high density.

\subsubsection{The remaining REBELS sources}
The non-detection of [OIII]$_{88\mu\text{m}}$ in REBELS-37 can be explained by its spectroscopic redshift. The peak of the redshift likelihood distribution shifted to $z<8$ after additional \textit{HST} observations (GO 16879, PI: Stefanon and GO 15931, PI: Bowler). Therefore, in Cycle 8 [CII]$_{158\mu\text{m}}$ was targeted instead and has been detected (\mbox{$z_{\mathrm{[CII]}} = 7.64$}, see Schouws et al. in preparation). Unfortunately, the Cycle 7 observations of REBELS-37 did not cover the [OIII]$_{88\mu\text{m}}$ emission line. 

For the other two sources, REBELS-11 and REBELS-13, the possibility that the [OIII]$_{88\mu\text{m}}$ line is at a frequency not covered by the SPWs of the line scans is substantial. Only $\sim$40-50\% of the redshift likelihood distributions of these two sources are covered by our observations. Additionally, $>$15$\%$ of the redshift likelihood distribution of REBELS-13 extends to $z < 6$, implying that this source could be at low redshift instead \citep{REBELS}. 

\subsection{Dust continuum emission from $z\gtrsim 7.6$ galaxies}
\label{sec:highz_dust}

Two of the four sources observed in Band 7, REBELS-04 and REBELS-37, show significant ($\geq$ 3.8$\sigma$) detections in the dust continuum. Interestingly, if the redshift of REBELS-04 is indeed $z \sim 8.6$ as its photometric redshift indicates, it would be the most distant dust continuum detected galaxy with ALMA. At present, the highest redshift source with a dust-continuum detection confirmed by spectroscopy is the $z=8.31$ \citet{tamura2019} source \citep{bakx2020}.  REBELS-04 would therefore be a compelling source to follow up with \textit{JWST} or with an ALMA Band 6 [CII]$_{158\mu\text{m}}$ spectral scan.

For REBELS-37 we have observations in both Band 6 and 7 and a spectroscopically confirmed redshift ($z_{\text{[CII]}} = 7.64$). Using the observations of the dust continuum in two ALMA bands, we have attempted to constrain the temperature of the dust in Section \ref{sec:tdust}. However, we do not have a detection in Band 6 and therefore only an upper limit that makes it significantly harder to get tight constraints on the dust temperature. With these measurements we can exclude low $T_{\text{dust}}$ ($\lesssim 28$ K) at $3\sigma$ and $T_{\text{dust}}$ ($\lesssim 42$ K) at $2\sigma$ assuming $\beta_{\text{IR}} = 2.03$.  

The dust temperatures implied by these constraints are generally higher than three sources with published constraints from REBELS \citep{algera_oiii_cii,algera2024} which lie in the range 32 to 42 K. Considering a broader sample, \citet{Mitsuhashi2024} observe galaxies at $5<$~$z< 8$ to show dust temperature in the range 30 to 80 K, with a median of 42 K. The dust temperature we infer for REBELS-37 is likely higher (at $2\sigma$ confidence) than the median temperature derived by \citet{Mitsuhashi2024}. Given the relatively small number of galaxies with dust temperatures in excess of 50 K, further targeting of the dust continuum of the source at higher frequencies (closer to the dust peak emission) would be useful for better constraining its dust temperature and exploring how hot dust can become in galaxies at $z > $ 6.

\section{Conclusions}
\label{sec:conclusions}

We have presented the results from the ALMA Large Program REBELS targeting [OIII]$_{88\mu\text{m}}$ in four UV-bright $z \gtrsim 7.6$ galaxies. In general the [OIII]$_{88\mu \text{m}}$ line is brighter at $z >$ 8 than the [CII]$_{158\mu \text{m}}$ line, motivating its use for the line scans of the $z \gtrsim 7.6$ REBELS sources.
Spectral scans of the four sources cover the frequency range of 326.4 to 373.0 GHz ($z=8.10$-9.39). These scans reached an [OIII]$_{88\mu \text{m}}$ luminosity ($5\sigma$) of $\sim$ 7.6 $\times\ 10^8\ \text{L}_{\odot}$ (assuming a FWHM of 400 km s$^{-1}$) equivalent to 33 $\text{M}_{\odot}\ \text{yr}^{-1}$.

None of the four galaxies scanned for [OIII]$_{88\mu\text{m}}$ reveal a credible line detection. 
The non-detection of REBELS-04 is a surprise given that our line scans cover $\geq 92\%$ of the redshift likelihood distribution and its inferred UV+IR SFR (40 $\text{M}_{\odot}\ \text{yr}^{-1}$). The approximate $5\sigma$ upper limit we obtain on $L_{\text{[OIII]$_{88\mu\text{m}}$}}$ line is $\sim$ 4.2 $\times$ 10$^8$ $\text{L}_{\odot}$ and 8.1~$\times$~10$^8$ $\text{L}_{\odot}$ assuming a FWHM of 100 and 400 km s$^{-1}$, respectively, for the [OIII]$_{88\mu \text{m}}$ line, consistent with SFRs 
of 18 and 36 $\text{M}_{\odot}\ \text{yr}^{-1}$, respectively, using the \citet{harikane2020} calibration.

As such, the source would appear to be faint in [OIII]$_{88\mu\text{m}}$ line emission relative to its SFR and more consistent with the [OIII]$_{88\mu\text{m}}$ deficit of local galaxies if we assume a FWHM of 100 km s$^{-1}$. A faint [OIII]$_{88\mu\text{m}}$ luminosity can be induced by various ISM conditions, such as a high density, low ionization parameter or a low metallicity. As the dust continuum detection of REBELS-04 suggests significant prior metal-enrichment, it is more likely that a high density and/or a low ionization parameter suppress the [OIII]$_{88\mu\text{m}}$ emission.

In the case of REBELS-37, which was subsequently found to have a redshift of 7.643 based on the detection of [CII]$_{158\mu\text{m}}$ (Schouws et al. in preparation), the non-detection is a simple consequence of the source lying outside the redshift range of our [OIII]$_{88\mu\text{m}}$ line scan.  For the two final sources with [OIII]$_{88\mu\text{m}}$ scans (REBELS-11 and REBELS-13), it is unclear whether the depth of our line scan or redshift coverage is the issue.

Two of the REBELS sources targeted with line scans in Band 7 (REBELS-04 and REBELS-37) show
significant Band 7 continuum emission ($\geq 3.8 \sigma$). If our previously published photometric redshift estimate $z_{\mathrm{phot}} = 8.57^{+0.10}_{-0.09}$ (or $z_{\mathrm{phot}} = 8.43^{+0.10}_{-0.10}$ correcting for an additional CGM component) for REBELS-04 is accurate, this galaxy could well constitute the most distant dust-detected galaxy identified with ALMA to date. Note that this redshift estimate is based on the apparent presence of a Lyman break in the photometry and a contribution of line emission to the 4.5$\mu$m flux (see Appendix~\ref{sec:phot_redshift}).  The significance of the dust-continuum detection we see in REBELS-04 (3.8$\sigma$) is higher than previously found by \citet{Inami2022} for the source (3.4$\sigma$) using only Cycle 7 observations. For REBELS-37 we also have an upper limit on the Band 6 continuum flux, from which we can conclude that cold temperatures ($\lesssim$ 28 K) are excluded at $3\sigma$ when we assume $\beta_{\text{IR}} = 2.03$. With the limited information from Band 6 and 7 we cannot accurately constrain the dust temperature, but it is likely that this source has warm or hot dust. 

To make further progress, we need a better understanding of the [OIII]$_{88\mu\text{m}}$/[CII]$_{158\mu\text{m}}$ luminosity ratio of UV-bright sources like the REBELS sample, in order to properly assess the expected [OIII]$_{88\mu\text{m}}$ luminosities and increase the success rate of further [OIII]$_{88\mu\text{m}}$ ALMA observations. To confirm the spectroscopic redshift of the three REBELS targets without a line detection, \textit{JWST} observations or additional ALMA [OIII]$_{88\mu\text{m}}$ (or [CII]$_{158\mu\text{m}}$) line scans need to be performed, preferably after additional rest-UV data constrains the redshift likelihood distribution more tightly for REBELS-11 and REBELS-13. Additional emission lines should also be observed, in order to constrain the ISM conditions of sources in which the [OIII]$_{88\mu\text{m}}$ luminosity is potentially fainter than expected, such as the [OIII]$_{52 \mu \text{m}}$ line with ALMA or [OIII]$_{5007}$  with \textit{JWST} to measure the density.

%%%%%%%%%%%%%%%%%%%%%%%%%%%%%%%%%%%%%%%%%%%%%%%%%%

\section*{Acknowledgements}

We are thankful to the referee for their constructive feedback, which substantially improved our analysis.

This paper makes use of the following ALMA data:  ADS/JAO.ALMA\#2019.1.01634.L. ALMA is a partnership of ESO (representing its member states), NSF (USA) and NINS (Japan), together with NRC (Canada), MOST and ASIAA (Taiwan), and KASI (Republic of Korea), in cooperation with the Republic of Chile. The Joint ALMA Observatory is operated by ESO, AUI/NRAO and NAOJ. 

RJB acknowledges support from NWO grants 600.065.140.11.N211 (vrijcompetitie) and TOP grant TOP1.16.057. 
JAH acknowledges support from the ERC Consolidator Grant 101088676 (VOYAJ).
HSBA and HI acknowledge support from the NAOJ ALMA Scientific Research Grant Code 2021-19A.
MA is supported by FONDECYT grant number 1252054, and gratefully acknowledges support from ANID Basal Project FB210003 and ANID MILENIO NCN2024$\textunderscore$112.
RAAB acknowledges support from an STFC Ernest Rutherford Fellowship [grant number ST/T003596/1]. 
YF is supported by JSPS KAKENHI Grant Numbers JP22K21349 and JP23K13149. IDL acknowledges funding from the European Research Council (ERC) under the European Union's Horizon 2020 research and innovation program DustOrigin (ERC-2019- StG-851622) and from the Flemish Fund for Scientific Research (FWO-Vlaanderen) through the research project G0A1523N. MS acknowledges support from the European Research Commission Consolidator Grant 101088789 (SFEER), from the CIDEGENT/2021/059 grant by Generalitat Valenciana, and from project PID2023-149420NB-I00 funded by MICIU/AEI/10.13039/501100011033 and by ERDF/EU.

\section*{Data Availability}
All ALMA observations used in this work are publicly available in the ALMA Science Archive (\url{https://almascience.nrao.edu/aq/}). \textsc{mf3d} is publicly available on github (\url{https://github.com/pavesiriccardo/MF3D}).


\bibliographystyle{mnras}
\bibliography{references}

%%%%%%%%%%%%%%%%%%%%%%%%%%%%%%%%%%%%%%%%%%%%%%%%%%

%%%%%%%%%%%%%%%%% APPENDICES %%%%%%%%%%%%%%%%%%%%%

\appendix
\section{The photometric redshift of REBELS-04}
\label{sec:phot_redshift}
In Section~\ref{sec:coverage_z} we show the redshift likelihood distributions for the four sources discussed in this work. These redshift likelihood distributions were first presented in \citet{REBELS} and are a combination of the results from different redshift likelihood codes. Recent work by \citet{asada2024} using \textsc{eazy-py} \citep{brammer2008, Brammer_eazy-py_2021} find that the inclusion of absorption due to hydrogen in the CGM can have an impact on the photometric redshift estimates one derives for distant sources.
In particular, \citet{asada2024} find that photometric redshifts can be systematically overestimated by $\delta z = 0.20$ when only accounting for the standard IGM absorption \citep[e.g.][]{Inoue2014_IGM} and a possible CGM component is neglected. Given these concerns, we look into the impact this can have on the inferred redshift for REBELS-04. 

The photometric results of REBELS-04 are shown in Table~\ref{tab:phot_R4}. The photometric measurements were obtained from a combination of deep ground-based and \textit{Spitzer}/IRAC photometry. The optical data include Subaru Hyper Suprime-Cam (HSC) imaging in the $g, r, i, z$, and $y$ bands (\citealt{aihara2018, aihara2018b}), along with additional coverage in the $u^*$, $g, r, i, y$, and $z$ bands from the \textit{Wide} component of the Canada–France–Hawaii Telescope Legacy Survey (CFHTLS; \citealt{erben2009, hildebrandt2009}). Near-infrared (NIR) coverage is provided by the UKIRT Infrared Deep Sky Survey (UKIDSS; \citealt{lawrence2007}) Ultra Deep Survey (UDS), complemented by the VISTA VIDEO survey (\citealt{jarvis2013}), which provides imaging in the $z$, $Y$, $H$, and $K_\mathrm{S}$ bands. The IRAC mosaics in the XMM-LSS field combine data from the Spitzer Large-Area Survey with HSC (SPLASH; \citealt{steinhardt2014}) and the Spitzer Extragalactic Representative Volume Survey (SERVS; \citealt{mauduit2012}).

Photometry across all bands was extracted using the \textsc{Mophongo} deblending procedure (\citealt{labbe2006, labbe2010a, labbe2010b, labbe2013, labbe2015}). The HSC $r$-band mosaic was used as a prior for source position and morphology, owing to its superior spatial resolution, as indicated by its PSF FWHM (see e.g. \citealt{stefanon2019}). Additionally, we incorporated \textit{HST}/WFC3 imaging in the F105W, F125W, and F160W bands from independent follow-up programs (PID 15931, PI: R. Bowler for F105W; PID 16879, PI: M. Stefanon for F125W and F160W). The observations consisted of 1-orbit exposures in the F105W band and 0.7–1.3 orbits in the F125W and F160W bands, respectively. The data were processed using a customized version of \textsc{MultiDrizzle} (\citealt{koekemoer2003}). Photometry in the \textit{HST} bands was extracted using \textsc{SExtractor} \citep{bertin1996} in dual-image mode, with detection performed on the F160W image. Flux densities were measured in $0\farcs6$-diameter apertures in each band and corrected for flux losses due to the finite aperture using the PSF curve of growth.

\begin{figure*}
\centering
    \includegraphics[width = \textwidth]{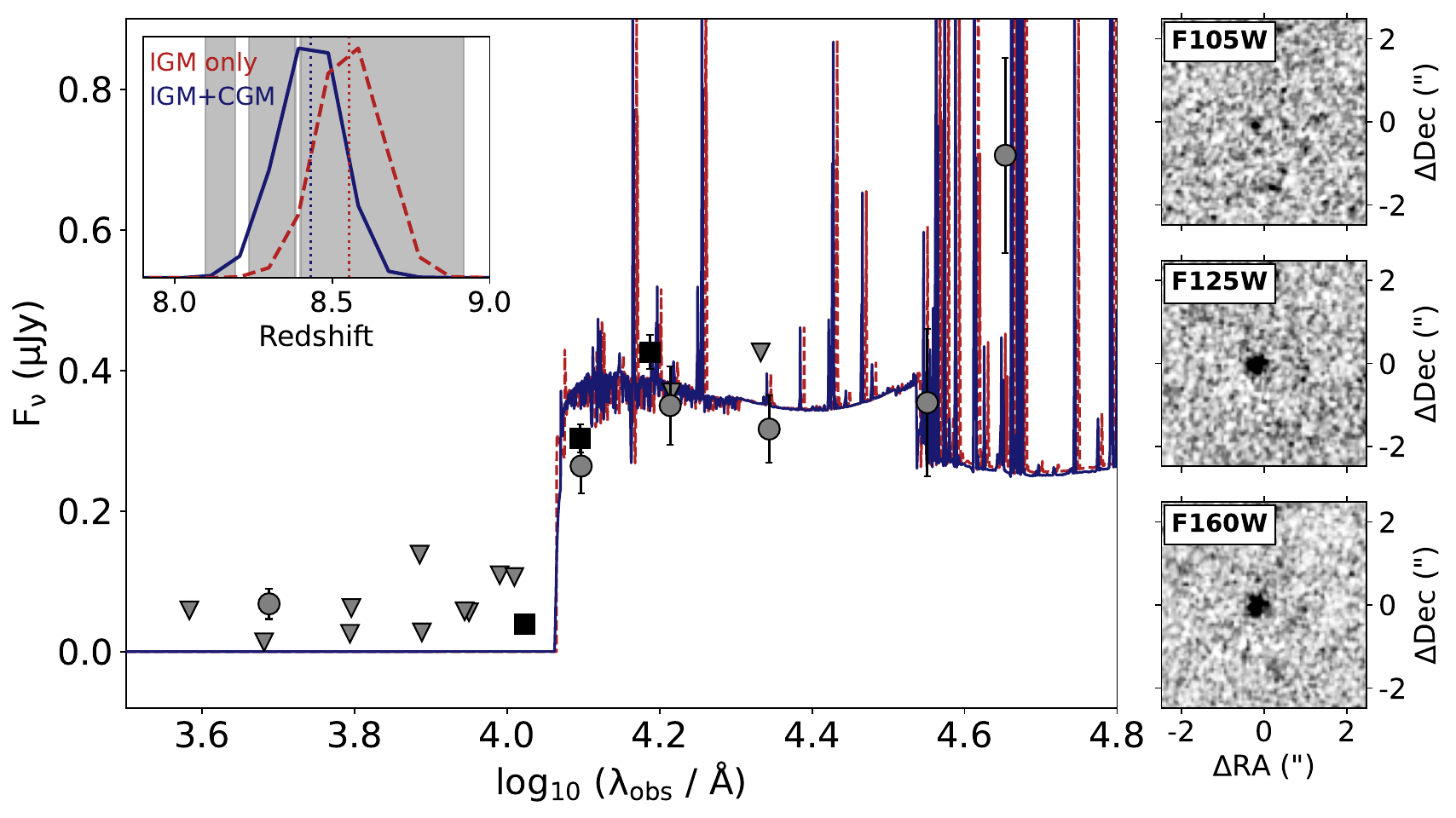}
    \caption{The rest-UV to optical SED for REBELS-04. The photometric measurements are shown with the circles and are from ground-based instruments and Spitzer. Non-detections ($<3\sigma$) are shown with $2\sigma$ upper limits (triangles), with the exception of the flux measured with CFHT $Z'$ that is outside of the range of the figure. The \textit{HST} observations in F105W, F125W and F160W are shown by the squares. The best-fit SED from \textsc{eazy-py} is plotted including the impact of the standard IGM transmission alone (\textit{dashed line}) and also including a CGM absorption component (\textit{solid line}). The gray area in the inset showing the redshift likelihood distributions shows the redshift range scanned for [OIII]$_{88\mu\text{m}}$ by Band 7 observations.  The addition of a CGM component as described by \citet{asada2024} decreases the median redshift (\textit{vertical dotted lines}) to $ z= 8.43^{+0.10}_{-0.10}$. The F105W, F125W and F160W imaging observations of REBELS-04 (4"$\times$4") with \textit{HST} are shown to the right of the SED and reveal a sharp Lyman break across the F105W and F125W filters.  Additionally, the \textit{Spitzer}/IRAC [3.6]-[4.5] color of REBELS-04 is 0.7$_{-0.1}^{+0.2}$ mag, strongly suggesting the presence of H$\beta$+[OIII]$_{4959,5007}$ line emission in the [4.5] band and a redshift in the range 7.0-9.1 \citep[e.g.][]{RobertsBorsani2016}.}
    \label{fig:sed_r04}
\end{figure*}

To examine the effect of CGM absorption on the photometric redshift we utilize \textsc{eazy-py} to fit the SED of REBELS-04. We perform two \textsc{eazy-py}, one including CGM and IGM absorption and one with only IGM absorption. The SED templates used in these two runs are from BPASS\footnote{https://bpass.auckland.ac.nz/}, as these were previously used to fit the SED of REBELS-04 and resulted the photometric redshift estimate (\mbox{$z=8.57$}) from \citet{REBELS}. The resulting best-fit templates and redshift likelihood distributions are shown in Figure~\ref{fig:sed_r04} by a solid and dashed line, respectively. The median redshift as shown by the dotted vertical line of the IGM only fit is $z = 8.55^{+0.11}_{-0.09}$, nearly identical to the $z = 8.57^{+0.10}_{-0.09}$ from \citet{bouwens2020} estimated using a similar approach.  For simplicity, we will quote the earlier photometric redshift estimate in representing this approach.
However, the addition of a CGM component decreases the redshift to $ z= 8.43^{+0.10}_{-0.10}$. This suggests that REBELS-04 could be at a slightly lower redshift than we had estimated earlier. The coverage from SPWs of the ALMA Band 7 observations of the redshift distribution from \textsc{eazy-py} is shown in Figure~\ref{fig:sed_r04} by the gray area. With the inclusion of CGM absorption the redshift likelihood distribution is covered by $92\%$ by the SPWs. Therefore, it seems very likely that the [OIII]$_{88\mu\text{m}}$ emission of REBELS-04 lies within the spectral scan range adopted for the source.

\begin{table}
    \centering
    \caption{The fluxes measured for REBELS-04 from both ground-based and space observations of rest-UV and optical wavelengths.}
    \label{tab:phot_R4}
    \begin{tabular}{|l|c|c|}
    \hline
    Telescope/Instrument & Filter & Flux (nJy) \\
    \hline
    \textit{HST}/WFC3 & F105W & 39 $\pm$ 10 \\
     & F125W & 304 $\pm$ 13 \\
     &F160W & 426 $\pm$ 12 \\ \hline
    CFHT/Megaprime & $U'$ & 10 $\pm$ 29 \\
     & $G'$ & 68 $\pm$ 21 \\
     &$R'$ & 10 $\pm$ 31 \\
     &$I'$ & -4 $\pm$ 69 \\
     &$Z'$ & 1073 $\pm$ 660 \\ \hline
    Subaru/HSC & $g$ & -6 $\pm$ 7 \\
     & $r$ & -2 $\pm$ 13 \\
     & $i$ & -14 $\pm$ 14 \\
     & $z$ & 60 $\pm$ 28 \\
     & $y$ & 2 $\pm$ 54 \\ \hline
    VISTA/VIRCAM & $Z$ & -34 $\pm$ 29 \\
     & $Y$ & -8 $\pm$ 53 \\
     & $H$ & 520 $\pm$ 182 \\
     & $K_s$ & 259 $\pm$ 212 \\ \hline
    UKIRT/WFCAM & $J$ & 264 $\pm$ 37 \\
      & $H$ & 350 $\pm$ 53 \\
      & $K$ & 317 $\pm$ 45 \\ \hline
    \textit{Spitzer}/IRAC & 3.6$\mu$m & 354 $\pm$ 103$^a$ \\
     & 4.5$\mu$m & 706 $\pm$ 134$^a$ \\
    \hline
    \end{tabular}\\
    \begin{flushleft}$^a$ Note that the \textit{Spitzer}/IRAC [3.6]-[4.5] color of REBELS-04 is 0.7$_{-0.1}^{+0.2}$~mag, strongly suggesting the presence of H$\beta$+[OIII]$_{4959,5007}$ line emission in the [4.5] band.  This provides further evidence that the source lies in the redshift range 7.0-9.1 \citep[e.g.][]{RobertsBorsani2016}.
\end{flushleft}
\end{table}

\section{Probable noise peak close to the rest-UV position of REBELS-13}
\label{sec:6787}

In Section \ref{sec:method_oiii} we explained the use of \textsc{mf3d} to identify line emission in our ALMA data cubes. Schouws et al. in preparation found a purity of $> 95\%$ for \mbox{SNR$_{\text{MF3D}} \geq 5.2$} emission peaks within a 1.5" radius of the center of the image. With \textsc{mf3d} we find a peak with SNR$_{\text{MF3D}}$ = 5.35 in the data cube of REBELS-13 1.45" from the center. However, upon further inspection of the spectrum corresponding to this peak and considering that it is very offset from the rest-UV position of REBELS-13, it is likely to be a noise peak. Observations of UV-bright sources have shown that the [OIII]$_{88\mu\text{m}}$ is often co-spatial with the UV emission (e.g. \citealt{harikane2020, witstok22}).

\begin{figure}
    \centering
    \includegraphics[width = 0.5\textwidth]{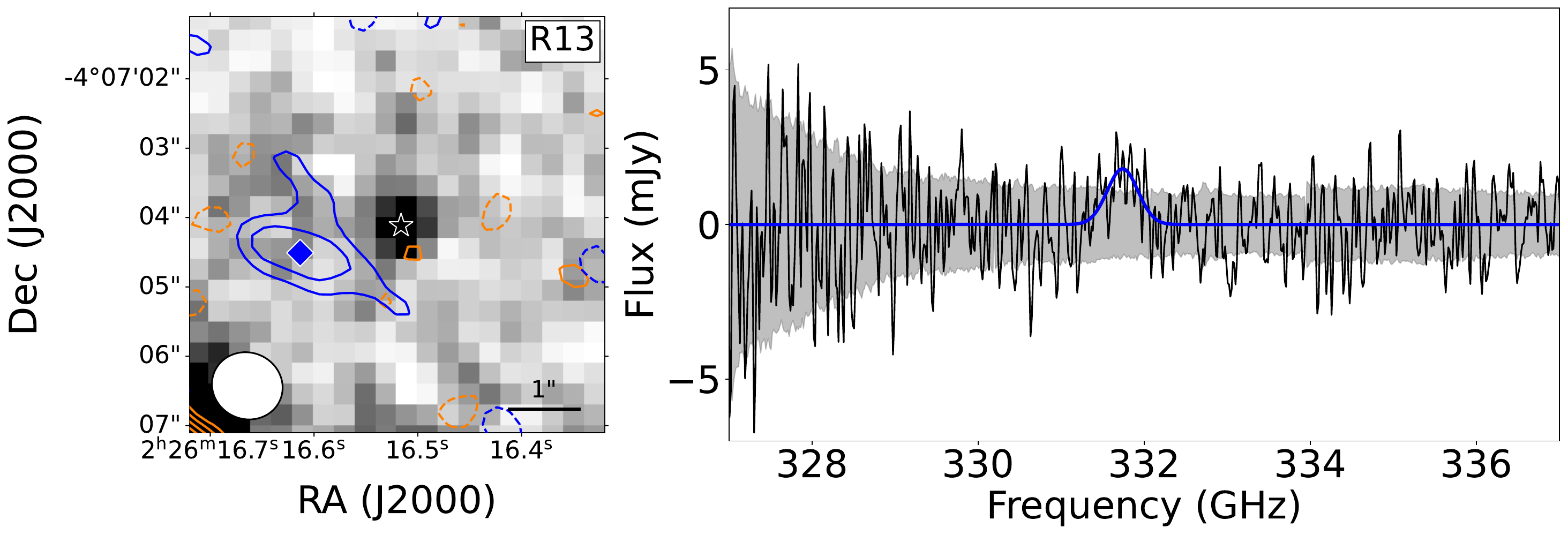}
    \caption{\textit{Left:} The SNR$_{\text{MF3D}} = $ 5.35$\sigma$ peak that is found 1.45" from the center of the observations of REBELS-13. The potential [OIII]$_{88\mu\text{m}}$ emission is plotted with blue contours and the corresponding beam of the moment-0 map is shown. The diamond indicates the peak emission in the moment-0 image, the star shows the peak of the rest-UV emission and the orange contours show the dust continuum emission. The contours run from 2$\sigma$ to 5$\sigma$ (in steps of 1$\sigma$) and negative contours are shown with dashed lines. \textit{Right:} The spectrum obtained with a circular aperture with 1" radius at the position of the peak with the 1$\sigma$ uncertainty determined from 1000 random apertures. Due to the low significance of the peak in the spectrum and the large offset from the rest-UV position we interpret it as a probable noise feature.}
    \label{fig:spec_13}
\end{figure}

In Figure \ref{fig:spec_13} we show the spectrum obtained by taking an aperture with a 1" radius at the position of the peak. The gray area shows the $1\sigma$ uncertainty on the spectrum calculated from 1000 random apertures, that shows that the significance of the line emission peak is $\lesssim 2 \sigma$. We fit a Gaussian to the spectrum by iteratively determining the FWHM of the line (starting from the frequency template from \textsc{mf3d}). We find that the line has a FWHM of 393$\pm 76$ km s$^{-1}$. The central frequency and FWHM of this fit are used to compute a moment-0 map including the frequencies within 2 $\times$ the FWHM using the \textit{immoments} task of \textsc{casa}. The resulting moment-0 map is also shown in Figure \ref{fig:spec_13} with contours from 2 to 5$\sigma$. To consider this offset peak as [OIII]$_{88\mu\text{m}}$ emission we would at least need to have a $5\sigma$ detection, which is not the case. Therefore, with our current observations we do not consider this peak for our analysis. 

\section{Dust SED fit of REBELS-37}
\label{sec:sed_fit}
REBELS-37 has been detected in Band 7 dust continuum emission ($\nu_\mathrm{obs}=349.7\,\mathrm{GHz}$), while later Band 6 observations (\mbox{$\nu_\mathrm{obs} = 219.2\,\mathrm{GHz}$}) did not yield a continuum detection. Given that dust continuum constraints are thus available at two distinct frequencies for REBELS-37, it is -- at least in theory -- possible to fit for its dust temperature and mass. We do so adopting the modified blackbody (MBB) fitting framework from \citet{algera_oiii_cii}, adopting optically thin dust emission and assuming a fixed dust emissivity index of either $\beta_\mathrm{IR}=1.5$ or $\beta_\mathrm{IR} = 2.0$. Following their work, we adopt a flat prior on the dust temperature between the CMB temperature at $z=7.643$ and $150\,\mathrm{K}$. Towards higher temperatures, we smoothly decrease the prior following a Gaussian with a standard deviation of $\sigma =30\,\mathrm{K}$. With this prior, the dust temperature and mass can be robustly inferred, provided sufficiently deep dust continuum constraints are available, as demonstrated through fitting mock ALMA observations of high-redshift galaxies (\citealt{algera2024}; see also Sommovigo et al.\ in preparation). 

\begin{figure}
    \centering
    \includegraphics[width = 0.45 \textwidth, trim={0 1.0cm 0 0.5cm}]{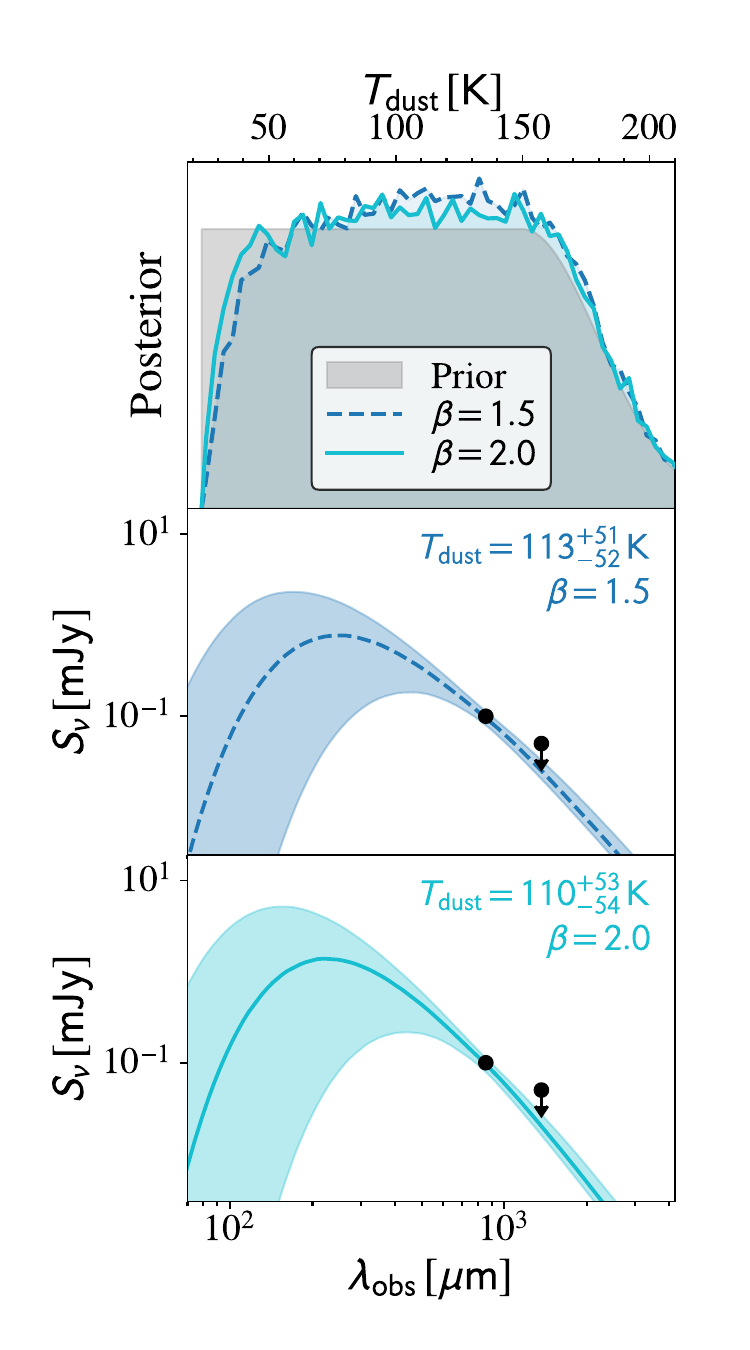}
    \caption{\textit{Top:} dust temperature posteriors obtained from fitting an optically thin MBB to the ALMA continuum observations of REBELS-37. A fixed value of $\beta_\mathrm{IR}=1.5$ (dashed blue line) or $2.0$ is assumed (solid cyan line). The adopted prior is moreover shown through the gray shading. \textit{Middle:} optically thin MBB fit with a fixed $\beta_\mathrm{IR}=1.5$. \textit{Bottom:} same as the middle panel, now using $\beta_\mathrm{IR}=2.0$. With the current dust continuum measurements, it is not possible to accurately constrain the dust temperature of REBELS-37. Only very cold temperatures ($T_\mathrm{dust} \lesssim 30\,\mathrm{K}$) can be ruled out with a high degree of confidence (see also Section \ref{sec:highz_dust}). Otherwise, the dust temperature -- and thus the dust mass -- of REBELS-37 remains nearly fully unconstrained. In practice, this means that the posterior distribution for the dust temperature resembles the prior, as can be seen in the top panel.}
    \label{fig:MBBfit}
\end{figure}

The two dust SED fits are shown in the middle and bottom panels of Figure \ref{fig:MBBfit}. The top panel shows the inferred posterior distribution of the dust temperature, for both values of $\beta_\mathrm{IR}$. While very cold dust temperatures can be ruled out for REBELS-37 ($T_\mathrm{dust}\lesssim30\,\mathrm{K}$; see also Section \ref{sec:highz_dust}), it is clear from the broad dust temperature posterior that the precise dust temperature cannot be constrained, irrespective of the assumed $\beta_\mathrm{IR}$. In practice, the posterior indeed resembles the prior for temperatures $T_\mathrm{dust} \gtrsim 50\,\mathrm{K}$. To better constrain the dust temperature and dust mass of REBELS-37, deeper observations are necessary -- ideally probing both the peak and Rayleigh-Jeans tail of the dust SED (e.g., \citealt{bakx2021,algera2024}). 

As the dust temperature of REBELS-37 cannot be robustly constrained with present data, we report a $3\sigma$ lower limit in Section \ref{sec:highz_dust} based on an analysis of the Band 6 and 7 continuum flux ratio. This yields a dust temperature of $T_\mathrm{dust} \gtrsim 28\,\mathrm{K}$ for REBELS-37.

%%%%%%%%%%%%%%%%%%%%%%%%%%%%%%%%%%%%%%%%%%%%%%%%%%

% Don't change these lines
\bsp	% typesetting comment
\label{lastpage}
\end{document}